\documentclass[10pt,conference]{IEEEtran}

\pdfoutput=1

\usepackage[noadjust]{cite}
\usepackage{graphicx}
\usepackage[cmex10]{amsmath}
\usepackage{amssymb}

\usepackage{array}

\usepackage{booktabs}
\usepackage{enumerate}
\usepackage[caption=false,font=footnotesize]{subfig}

\begin{document}

\title{Randomized Last-Level Caches Are Still Vulnerable to Cache Side-Channel Attacks! But We Can Fix It}

\author{
  \IEEEauthorblockN{
    Wei~Song\IEEEauthorrefmark{1}\IEEEauthorrefmark{2},
    Boya Li\IEEEauthorrefmark{1}\IEEEauthorrefmark{2},
    Zihan Xue\IEEEauthorrefmark{1}\IEEEauthorrefmark{2},
    Zhenzhen Li\IEEEauthorrefmark{1}\IEEEauthorrefmark{2},
    Wenhao Wang\IEEEauthorrefmark{1}\IEEEauthorrefmark{2},
    Peng Liu\IEEEauthorrefmark{3}}
  \IEEEauthorblockA{\IEEEauthorrefmark{1}State Key Laboratory of Information Security, Institute of Information Engineering, CAS, Beijing, China}
  \IEEEauthorblockA{\IEEEauthorrefmark{2}School of Cyber Security, University of Chinese Academy of Sciences, Beijing, China}
  \IEEEauthorblockA{\IEEEauthorrefmark{3}The Pennsylvania State University, University Park, USA}
  \IEEEauthorblockA{\{songwei, liboya, xuezihan, lizhenzhen1, wangwenhao\}@iie.ac.cn, pxl20@psu.edu}
}

\maketitle

\begin{abstract}
  Cache randomization has recently been revived as a promising defense against conflict-based cache side-channel attacks.
  As two of the latest implementations,
  CEASER-S and ScatterCache both claim to thwart conflict-based cache side-channel attacks using randomized skewed caches.
  Unfortunately,
  our experiments show that an attacker can easily find a usable eviction set within the chosen remap period of CEASER-S
  and increasing the number of partitions without dynamic remapping, such as ScatterCache, cannot eliminate the threat.
  By quantitatively analyzing the access patterns left by various attacks in the LLC,
  we have newly discovered several problems with the hypotheses and implementations of randomized caches,
  which are also overlooked by the research on conflict-based cache side-channel attack.

  However, cache randomization is not a false hope and
  it is an effective defense that should be widely adopted in future processors.
  The newly discovered problems are corresponding to flaws
  associated with the existing implementation of cache randomization and are fixable.
  Several new defense techniques are proposed in this paper.
  our experiments show that all the newly discovered vulnerabilities of existing randomized caches
  are fixed within the current performance budget.
  We also argue that randomized set-associative caches can be sufficiently strengthened and
  possess a better chance to be actually adopted in commercial processors than their skewed counterparts
  as they introduce less overhual to the existing cache structure.

\end{abstract}

\section{Introduction}

To reduce the latency of accessing memory,
modern computers adopt a multi-level cache hierarchy
where the last-level cache (LLC) is shared between all processing cores.
Such sharing improves the utilization efficiency of the LLC
as it can dynamically adapt its space allocation to the demand
of different cores.
However, it also allows a malicious software to
trigger controlled conflicts in the LLC, such as evicting a specific cache set with attackers' data~\cite{Percival2005, Osvik2006, Liu2015},
to infer security-critical information of a victim program.
This type of conflict-based cache side-channel attacks have been utilized to recover cryptographic keys~\cite{Irazoqui2015},
break the sandbox defense~\cite{Genkin2018},
inject faults directly into the DRAM~\cite{Gruss2016a},
and extract information from the supposedly secure SGX enclaves~\cite{Haehnel2017}.

Cache partitioning~\cite{Page2005, Kim2012,Liu2016} used to be
the only effective defense against conflict-based cache side-channel attacks abusing the LLC.
It separates security-critical data from normal data in the LLC; therefore,
attackers cannot evict security-critical data by triggering conflicts using normal data.
However, cache partitioning is ineffective when security-critical data cannot be
easily separated from normal data~\cite{Gruss2016b}
or normal data become the target~\cite{Gruss2016a}.
It also reduces the autonomy of the LLC which in turn
hurts performance.
Finally, cache partitioning relies on specific operating system (OS) code to identify security-critical data,
which means the OS must be trusted.

Recently, cache randomization~\cite{Wang2007, Wang2008, Liu2014, Qureshi2018, Qureshi2019, Werner2019, Ramkrishnan2019, Doblas2020, Tan2020}
has been revived as a promising defense.
Instead of cache partitioning,
cache randomization randomizes the mapping from memory addresses to cache set indices.
This forces attackers to slowly find eviction sets using search algorithms at run-time~\cite{Oren2015, Liu2015, Vila2019, Song2019}
rather than directly calculating eviction set indices beforehand.
Even when eviction sets are found, attackers cannot tell which cache sets are evicted by them.
However, cache randomization alone does not defeat conflict-based cache side-channel attacks
but only increases difficulty and latency~\cite{Qureshi2018}.
For this reason,
dynamic remapping~\cite{Qureshi2018, Ramkrishnan2019} has been introduced to limit the time window available to attackers
and skewed cache~\cite{Qureshi2019, Werner2019, Ramkrishnan2019} has been proposed to further increase the attack difficulty.

As two of the latest implementations,
CEASER-S~\cite{Qureshi2019} and ScatterCache~\cite{Werner2019} both claim to thwart conflict-based cache side-channel attacks
using randomized skewed caches.
ScatterCache even argues that dynamic remapping might not be necessary as the extra difficulty introduced by skewed cache is hard enough.

Unfortunately,
our experiments show that an attacker can easily find a usable eviction set within the chosen remap period of CEASER-S~\cite{Qureshi2019}
and increasing the number of partitions without dynamic remapping, such as ScatterCache~\cite{Werner2019}, cannot eliminate the threat.
By quantitatively analyzing the access patterns left by various attacks in the LLC,
we have newly discovered several problems with the hypotheses and implementations of randomized caches,
which are also overlooked by the research on conflict-based cache side-channel attacks.

\begin{itemize}
\item \emph{The possibility of using cache flush instructions in conflict-based attacks has been overlooked.}
  Our study shows, if attackers flush the eviction set after each probe,
  partial congruent eviction sets can be repeatedly used to drastically speed up attacks
  and significantly reduce the latency in finding eviction sets.
\item \emph{The concept of minimal eviction set no longer applies to randomized skewed caches.}
  Any group of cache blocks that can evict the target address with a reasonable probability
  should be considered a usable eviction set.
\item \emph{Attackers do not have to use eviction sets with 99\% eviction rate.}
  When finding such sets become too difficult,
  attackers will utilize eviction sets with low eviction rate but possible to find.
\item \emph{Measuring the remap period by LLC accesses is flawed,}
  since a significant portion of all the cache accesses might be filtered by the private level-one (L1) or level-two (L2) caches.
  The actual number of accesses observed by the LLC is much smaller than the total number of cache accesses.
  As a result, the remap period estimated by CEASER-S~\cite{Qureshi2019} is over-optimistic.
\end{itemize}

However, cache randomization is not a false hope.
We strongly believe it is an effective defense strategy that should be widely adopted in future processors.
The above-discovered problems are corresponding to flaws
associated with the existing tactics towards accomplishing the ``cache randomization'' strategy.
We believe that these problems are fixable,
and that fixing these problems will make the strategy significantly more effective in defending conflict-based cache side-channel attacks.
In particular, several new defense techniques/suggestions are proposed in this paper:

\begin{itemize}
\item \emph{Measure the remap period by LLC evictions rather than accesses}
  because the probability of successfully finding an eviction set is closely related to
  the number of evictions allowed between remaps.
\item \emph{Further reduce the period} to stop attackers from finding even small partially congruent eviction sets.
\item \emph{Adopte ZCache-like~\cite{Sanchez2010} multi-step relocation to minimize the number of cache blocks evicted during the remap process.}
\item \emph{Promote the use of CEASER (randomized set-associative cache) rather than skewed caches}
  because CEASER introduces less overhual to the existing cache structure than skewed caches and it can be made secure enough.
\item \emph{A simple attack detection mechanism to further strengthen CEASER.}
\end{itemize}

By utilizing these defense techniques/suggestions,
our experiments show that all the newly discovered vulnerabilities of existing randomized caches can be fixed within the current performance budget
and the randomized set-associative caches can be made secure enough with reasonable performance overhead.

This paper is organized as follows:
Section~\ref{sec:background} introduces the necessary background information to understand this paper.
Section~\ref{sec:problem} formulates the problems we try to answer in this paper.
Section~\ref{sec:vulnerable} demonstrates the vulnerabilities of existing randomized caches by experiments.
Section~\ref{sec:fix-skew} shows how we can fix the randomized skewed caches
and Section~\ref{sec:fix-ceaser} presents solutions to safely strengthen the randomized set-associative caches.
The performance overhead is analyzed in Section~\ref{sec:perf}.
The limitations and related work are discussed in Section~\ref{sec:related}.
Section~\ref{sec:con} finally concludes the paper.

\section{Background}\label{sec:background}

\subsection{Caches}

\begin{figure}[bt]
\centering{
\includegraphics[width=0.40\textwidth]{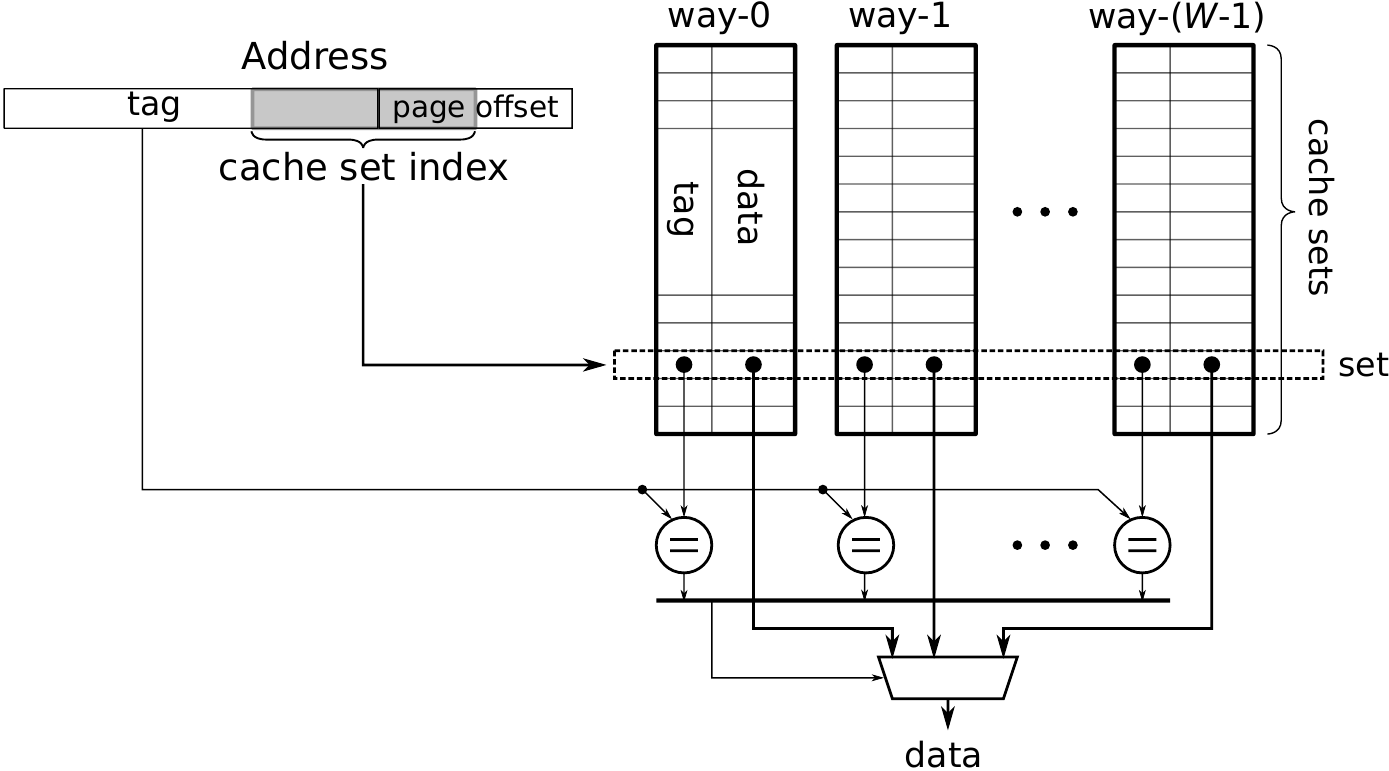}
}
\caption{
  A set-associative cache.
}
\label{fig:cache}
\end{figure}

Modern processors use caches to store recently or frequently used data to reduce the memory access time.
Most caches adopts a set-associative structure~\cite{Cekleov1997a} as shown in \figurename~\ref{fig:cache}.
The cache space is divided into $S$ cache sets
and each set contains $W$ ways of cache blocks.
Cache sets are addressed by a cache set index
which is typically a subset of the address bits
shared by all cache blocks in the same set.
If two addresses are mapped to the same cache set,
they are congruent addresses~\cite{Vila2019}.
When an address is accessed,
the cache checks whether there is a match (hit) in the corresponding cache set by comparing tags.
If no match is found (a miss), the cache block is fetched and stored in the cache set for future use.
The specific position (way) to store this newly fetched cache block is chosen by a replacement policy and the old block is evicted.
As a commonly used replacement policy, least-recently used (LRU)~\cite{Berg2006} retains the recently accessed cache blocks.

Multiple levels of caches are normally hierarchically organized.
A processing core might have one or two levels of private caches (L1 and L2 caches)
while all cores share a large LLC.
An inclusive relationship between private caches and the LLC is usually adopted~\cite{Cekleov1997a}.
When a cache block is evicted from the LLC, it is also purged from all private caches.
A hardware managed coherence protocol ensures data are correctly updated between caches.

\subsection{Conflict-Based Cache Side-Channel Attacks}

Conflict-based cache side-channel attacks~\cite{Yan2017} exploit the fact that
cache blocks in the same set are congruent.
This allows attackers to maliciously control the status of a target cache set
using a group of at least $W$ congruent addresses (cache blocks), namely an eviction set.

An attack normally occurs in two phases:
\emph{preparation phase} when the attacker collects enough number of eviction sets,
and \emph{exploitation phase} when the attacker infers sensitive information from a victim
by controlling the status of certain cache sets using the collected eviction sets.

Before cache randomization is applied,
collecting eviction sets are relatively easy
because attackers can deliberately construct an eviction set
using addresses having the same cache set index bits~\cite{Liu2015}.
This becomes unfeasible when caches are randomized.

The exploitation phase normally contains numerous prime+probe cycles~\cite{Percival2005, Osvik2006, Liu2015}.
In each cycle, the attacker first \emph{primes} a target set by filling it with
cache blocks from a corresponding eviction set.
If there were cache blocks belonging to the victim, they are likely evicted in the \emph{prime} process.
The attacker then tricks the victim into running a program segment related to the target cache set.
If the victim indeed accesses data indexed to the same cache set,
it must have been fetched into the cache set and one block of the eviction set is consequently evicted.
Finally the attacker \emph{probes} the cache set by re-accessing all blocks of the eviction set.
If the total access latency is longer than expected,
the attacker learns that the victim should have accessed the target cache set,
which might further infer other security-critical information.

\subsection{Randomized Caches}\label{sec:random-cache}

\begin{figure}[bt]
\centering{
\includegraphics[width=0.40\textwidth]{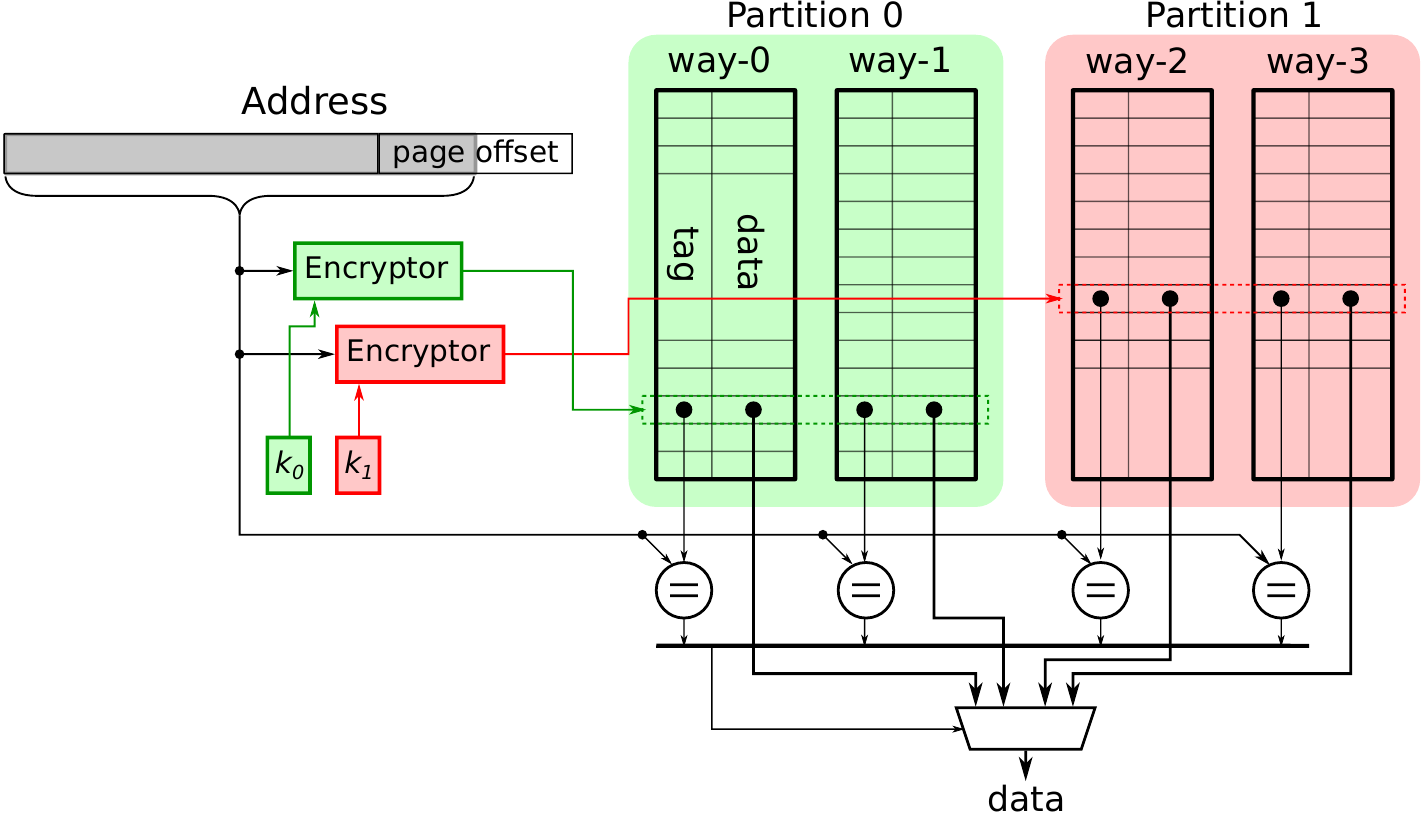}
}
\caption{
  A randomized skewed cache with two partitions over four cache ways.
}
\label{fig:skewed-cache}
\end{figure}

The main objective of cache randomization is to deprive attackers from useful eviction sets~\cite{Qureshi2018, Qureshi2019, Werner2019}.
The latest implementation of cache randomization is randomized skewed caches~\cite{Qureshi2019, Werner2019},
while randomized set-associative caches~\cite{Qureshi2018} can be considered as a special case with only one partition.
\figurename~\ref{fig:skewed-cache} presents a randomized skewed cache 
whose four cache ways are evenly divided into two partitions independently indexed.
Instead of using a subset of address bits,
the cache set index is generated from an encryptor taking the whole address and a hardware managed key as inputs.
Assuming the encryption algorithm is unbroken and the key is not leaked,
the cache set index is a random number unobservable to attackers.
They can no longer construct eviction sets simply by picking addresses
but dynamically search for congruent addresses through run-time experiments,
which was considered an intolerable long procedure~\cite{Liu2015, Oren2015, Vila2019, Song2019}.

Another major benefit of randomized skewed caches is the reduced effectiveness of eviction sets.
Considering two random addresses,
they are fully congruent when they are mapped to the same sets in all partitions
while they are partially congruent when they are mapped to the same sets in some but not all partitions.
A group of $W$ addresses, where $W$ is the number of ways, forms a fully congruent eviction set
only when all of the $W$ addresses are fully congruent.
However, the probability that two random addresses are fully congruent in a $K$ partitioned skewed cache
is $\frac{1}{S^K}$, where $S$ is the number of cache sets.
This is an extremely small probability when $K$ is large.
Finding such a fully congruent eviction set at run-time is unfeasible.
Therefore, attackers have no choice but to use partially congruent eviction sets composed of partially congruent addresses.
This has two drawbacks~\cite{Werner2019, Purnal2019}:
The number of addresses needed is significantly increased
and the eviction of the target address becomes a statistically random event.

\subsection{Fast Algorithms for Searching Eviction Sets}\label{sec:algorithm}

At the time when CEASER was proposed,
the fastest algorithm~\cite{Liu2015, Oren2015} for finding a minimal eviction set with $W$ addresses
required $\mathcal{O}(N^2)$ cache accesses,
where $N$ is the number of addresses randomly collected to form a very large eviction set.
As $N$ is normally at the magnitude with the size of the LLC ($N \sim S \cdot W$) ~\cite{Song2019},
$\mathcal{O}(N^2)$ cache accesses are just too long for any practical attacks.
Soon afterwards,
three fast search algorithms are proposed to drastically reduce the accesses.

\begin{figure}[bt]
\centering{
\includegraphics[width=0.35\textwidth]{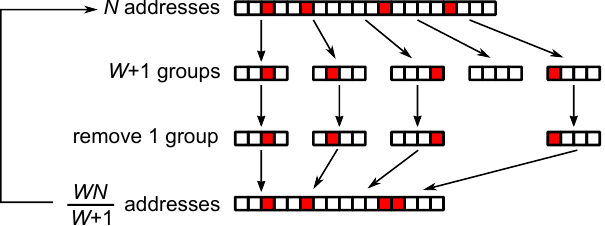}
}
\caption{
  Group elimination algorithm.
}
\label{fig:group-elimination}
\end{figure}

\emph{Group elimination} (GE) is an optimization of the original $\mathcal{O}(N^2)$ method~\cite{Vila2019, Qureshi2019}.
It still starts with a very large eviction set of $N$ random addresses
but it tries to remove multiple addresses in each cycle to quickly shrink the set into a minimal one.
\figurename~\ref{fig:group-elimination} illustrates such a cycle targeting an LLC with four ways.
The set of $N$ addresses are divided into $W+1$ groups.
Since a minimal eviction set contains only $W$ addresses (shadowed in red),
there is at least one removable group containing none of the $W$ addresses.
By sequentially testing whether the set is still an eviction set without a certain group,
the removable group is found and removed.
Then the whole process starts again taking the remaining $\frac{WN}{W+1}$ addresses as the input set
until a minimal set is produced.
The whole process requires around $\mathcal{O}(WN)$ cache accesses.
Since $N \sim SW$, $\mathcal{O}(WN) \sim \mathcal{O}(SW^2)$.

\emph{Conflict testing} (CT) is a new algorithm first proposed to
find eviction sets in caches using random replacement~\cite{Qureshi2019}.
Assuming an attacker has access to unlimited number of random addresses,
she can collect an eviction set by sequentially testing each address
whether it is congruent with the target address.
The target address is accessed first to make it cached in the LLC.
Then a random address is accessed.
If this address is congruent with the target address,
it might replace the target address by a chance of $\frac{1}{W}$ thanks to the random replacement.
Overall, any random address might conflict with the target address by a probability of $\frac{1}{S \cdot W}$.
To test the occurrence of such a conflict, the target address is re-accessed and timed.
If the latency is longer than expected, the random address is considered congruent and put into the eviction set.
The re-accessing of the target address also starts the test for the next random address.
An eviction set is produced when enough congruent addresses are collected.
The overall number of cache accesses is estimated around $\mathcal{O}(SW^2)$.

Note that this algorithm is effective for LLCs using permutation-based replacement (such as LRU) as well.
Assuming the use of LRU,
the probability of causing a conflict with the target address after accessing $M$ random addresses
is around:
\begin{equation}\label{eqn:conflict-lru}
P = 1 - \sum_{i=0}^{W-1}\binom{M}{i}\frac{1}{S^i}(1-\frac{1}{S})^{M-i}
\end{equation}
This is equivalent to causing at least $W$ conflicts in the target cache set.
The average $\overline{M}$ is around $SW$.
Note that re-accessing the target address is unlikely to cause an actual access to the LLC
because the target address is always cached in private caches (L1)
until it is forcefully evicted by a conflict in the LLC.
As a result, the LRU replacer's internal state is unchanged for most re-accessing of the target address.
To find a minimal eviction set with $W$ addresses,
the number of cache accesses is also around $\mathcal{O}(SW^2)$.

\begin{figure}[bt]
\centering{
\subfloat[Internal states]
         {\includegraphics[width=0.13\textwidth]{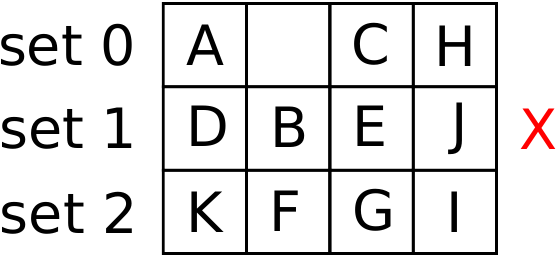}\label{fig:ppt-cache-state}}\\
\subfloat[Ideal case]
         {\includegraphics[width=0.27\textwidth]{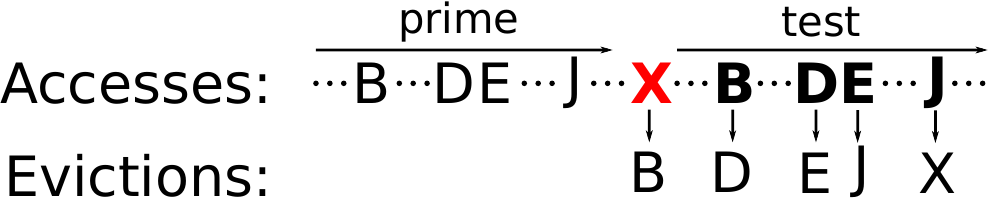}\label{fig:ppt-ideal}} \\
\subfloat[Non-ideal case]
         {\includegraphics[width=0.36\textwidth]{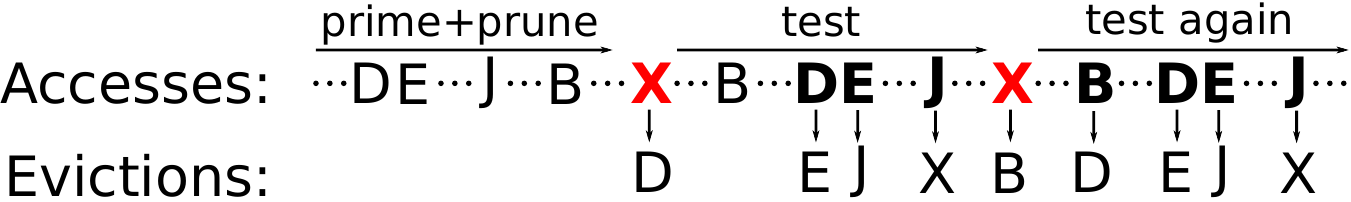}\label{fig:ppt-non-ideal}} \\
}
\caption{
  Prime, prune and then test on a 3-set 4-way LLC using the LRU replacement.
  (a) reveals the LLC internal states after prime and prune.
  (b) demonstrates a success test in an ideal case where all cache accesses are observed by the LLC.
  (c) shows a partial result in a non-ideal scenario
  when some cache accesses are filtered by the private caches
  and the observed orders of prime+prune and test are different.
}
\label{fig:prime-prune-test}
\end{figure}

\emph{Prime, prune and then test} (PPT) is an improved version of
the search algorithm exploiting the LRU replacement~\cite{Qureshi2019, Purnal2019}.
Let us consider an LLC using the LRU replacement.
An attacker first accesses a large set of random addresses (prime set) to prime the whole LLC\footnote{
  The attacker might choose to use a small set to prime a portion of the LLC
  but this will significantly reduce the success rate.
}.
Since self-conflicts would naturally occur during the prime,
a prune process is used to remove conflicted addresses
until all addresses remaining in the prime set are concurrently cached.
Assuming \figurename~\ref{fig:ppt-cache-state} reveals the internal states of an LLC after prime and prune,
the target cache set (set 1) is likely primed by the prime set.
In an ideal scenario, the order of cache accesses observed by the LLC is the same order initiated by the attacker.
As shown in \figurename~\ref{fig:ppt-ideal}, if the attacker makes a timed re-access of the target address X and the prime set sequentially,
all the addresses with long latency (miss in the LLC) are congruent
and the number of them is just enough for an eviction set.
However, the order seen by the LLC is normally different from the software order as many cache accesses are filtered by the private caches.
In this scenario (\figurename~\ref{fig:ppt-non-ideal}),
the attacker collects some but less than $W$ congruent addresses.
Normally she just has to test again to force the order.
As for the total number of cache accesses,
our experiments show that the prune process normally finishes in less than two rounds.
Meanwhile, the the size of the prime set after pruning is slightly less than the cache size,
which means only one round of search is usually enough.
The overall number of cache accesses is estimated around $\mathcal{O}(SW)$, which is the smallest in the three fast algorithms.

This algorithm can be used to find eviction sets in LLCs using other types of replacement policies.
The estimated number of accesses for permutation-based replacement policies (including LRU) is normally the same ($\mathcal{O}(SW)$)~\cite{Qureshi2019}
but it approaches to $\mathcal{O}(SW^2)$ for LLCs using random replacement for two reasons:
One is the size of the prime set after pruning is much smaller than the cache size $SW$,
which reduces the chance of finding congruent addresses in each round of search.
The other one is that, even if the target cache set is primed,
the number of congruent addresses found in each round of test is significantly less than $W$ (only one in most cases).
The attacker has to do multiple rounds of tests in multiple rounds of searches~\cite{Purnal2019}.

\subsection{Attack Randomized Caches using the Fast Algorithms}

All the three search algorithms can easily defeat the static version of randomized caches, such as CEASE~\cite{Qureshi2018}.
As a result, a randomized cache has to periodically remap its content
by updating the hardware managed key (\figurename~\ref{fig:skewed-cache}).
This forces an attacker to dynamically search eviction sets and
finish an attack both in the remap period.
Short remap period increases the hardness to launch an attack~\cite{Qureshi2018}.

However, frequent remaps lead to significant performance loss.
During the remap process,
all cache blocks in the LLC are sequentially relocated using the updated key.
When there is no available space at the new location,
a cache block is evicted to make space~\cite{Qureshi2018}.
Our experiments show that 40\% to 50\% cache blocks
are evicted for this reason.

To reduce the performance overhead while thwarting attacks,
the remap period is carefully selected.
For a 1024-set 16-way CEASER LLC,
it has to remap around every 47K accesses (only three accesses per cache block)~\cite{Qureshi2019},
which is an unbearably short period.
This is why skewed caches are currently preferred.
For a same sized CEASER-S LLC with two partitions,
it is claimed that the remap period can be safely increased to 1.6M accesses (100 accesses per cache block)~\cite{Qureshi2019}.

It becomes almost impossible to find (fully congruent) eviction sets
in a randomized skewed cache remapped at the aforementioned rate.
In its current form,
the \emph{group elimination} algorithm simply fails in skewed cache
due to the huge amount of false negative errors introduced by the randomly selected partitions.
Both the \emph{conflict testing} and the \emph{prime, prune and then test} algorithms
might still be able to find partially congruent eviction sets~\cite{Purnal2019},
which is the root of concern found in this paper.

\section{Problem Formulation}\label{sec:problem}

In this paper, we would like to thoroughly examine the effectiveness of cache randomization.
To be specific, we plan to answer the following questions.

\textbf{Problem Statement:}
\begin{itemize}
\item Do the existing cache randomization schemes/techniques make any flawed hypothesis?
\item If there are any flawed hypothesis, do they lead to broken defenses and discovery of new vulnerabilities?
\item Whether the broken defenses, if any, can be fixed?
\end{itemize}

Before diving into the detailed analysis,
let us first describe the threat model and the analysis platform.

\subsection{Threat Model}

The objective of using cache randomization is to deprive attackers from useful eviction sets.
We thus consider finding a usable eviction set targeting a specific address as a successful attack.
Only conflict-based cache side-channel attacks targeting the LLC is considered in this paper.
We assume the attacker has the following favorable but still reasonable capabilities:

\begin{itemize}
\item She has fully reverse engineered the virtual to physical address mapping.
\item She has access to unlimited number of random addresses.
\item She can make arbitrary memory access to her own data and
  accurately infer cache hit/miss status by measuring the access latency.
\item She can flush a cache block from the whole cache hierarchy as long as it is her own data.
\item She can accurately trick the victim into running a single memory access and
  there is no other active process during the attack.
\item She has the full design details of the randomized cache
  but the encryption algorithm and the key used for generating cache set indices
  are unbreakable (we do not consider attacks targeting weak encryptors~\cite{Bodduna2020} or random number generators).
\end{itemize}

Note that we have explicitly allowed the attacker to flush her own data
and this is different from flush-reload attacks~\cite{Yarom2014} because there is no shared data between the attacker and the victim.
It is normally not required for conflict-based attacks targeting non-randomized caches
but attackers do have such capability, such as a malicious user mode program running on a x86-64 processor~\cite{Intel2015}
or an malicious kernel running on an ARM processor~\cite{Zhang2016a}.
As described in Section~\ref{sec:flaw-hypo},
allowing this enables attackers to launch attacks on the latest randomized caches using partially congruent eviction sets.

\subsection{Analysis Platform}\label{sec:ana-platform}

To quantitatively analyze the effectiveness of the latest randomized caches,
we choose to implement CEASER-S and ScatterCache in a behavioral cache simulation model opensourced by \cite{Song2019},
further extend the model with the defense techniques newly proposed in Section~\ref{sec:fix-skew} and \ref{sec:fix-ceaser},
and attack the randomized caches using the aforementioned fast search algorithms.
All results revealed in Section~\ref{sec:vulnerable}, \ref{sec:fix-skew} and \ref{sec:fix-ceaser}
are obtained from these experimental attacks.
To evaluate the impact of the new defense techniques on normal applications in Section~\ref{sec:perf},
we run the SPEC CPU 2006 benchmark cases~\cite{Henning2006} on
the RISC-V~\cite{Waterman2019} instruction level simulator Spike~\cite{Spike}
with its original cache model replaced with the extended cache model~\cite{Song2019}.
The use of Spike allows us to run benchmark cases at a speed around 1.5 million instructions per second,
which is ten times faster~\cite{Ta2018} than the Gem5 simulator~\cite{Binkert2011}
used in CEASER-S~\cite{Qureshi2019} and ScatterCache~\cite{Werner2019}.

\section{Dynamically Randomized Skewed Caches Are Still Vulnerable}\label{sec:vulnerable}

By quantitatively analyzing the traces left by various attacks in the LLC,
this section reveals the flawed hypotheses found in the existing randomized caches
and uses experimental attacks to show that the defenses are indeed broken.

\subsection{Flawed Hypothesis}\label{sec:flaw-hypo}

\begin{figure}[bt]
\centering{
\includegraphics[width=0.23\textwidth]{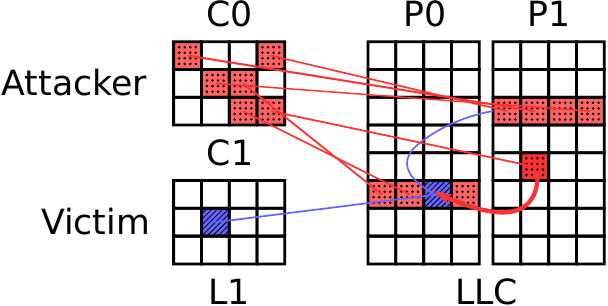}
}
\caption{
  When all cache blocks of the partially congruent eviction set (dot with red shadow) are cached in the LLC
  but failed to evict the target block (slash line with blue shadow),
  the eviction set become useless.
}
\label{fig:hard-evset}
\end{figure}

\textbf{Flawed hypothesis in CEASR-S}:
It is claimed in \cite{Qureshi2019} that
an attacker must find eviction sets composed of fully congruent cache blocks
in order to evict the target address repeatedly.
This is true for certain scenarios but not always true.
In order to illustrate why this is not always true, let us first reflect on a ``true'' scenario,
which is depicted in \figurename~\ref{fig:hard-evset}.
An attacker wants to launch a cross-core attack from core zero (C0) to core one (C1).
She has found an eviction set composed of seven fully congruent and one partially congruent cache blocks (dot with red shadow),
which should have a 50\% probability to evict the target address (slash line with blue shadow)
in a skewed cache with two partitions.
Assuming the the attacker has successfully evicted the target address several times,
she will fail eventually as the partially congruent cache block is randomly cached in the wrong partition (P1)
as depicted in \figurename~\ref{fig:hard-evset}.
The eviction set becomes useless afterwards.

From the attacker's viewpoint,
the reason of the failure is the lack of enough self-conflicts to
dislodge the misplaced partially congruent cache blocks during the re-accessing.
To reuse a eviction set, attackers must find another way to purge the misplaced blocks from the LLC.
Although one research~\cite{Purnal2019} claims that it is still viable to construct covert channels by priming the LLC,
this would cause significant amount of noise and noticeable performance degradation for normal prime-probe attacks.
We argue that \emph{attackers can accurately flush the eviction set
using cache flush instructions} (such as \texttt{clflush} in x86-64),
which is much cleaner and faster than priming the LLC.
Our argument indicates that attacks using partially congruent eviction sets could enjoy big success. 
Note that using flush instructions here is fundamentally different with the flush-reload attack~\cite{Yarom2014}
where the target address \emph{shared} between the attacker and the victim is flushed.
All blocks in an eviction set belong to the attacker's own address space.

We also argue this is a valid threat even for future computers
because \emph{the cache flush instructions will be here to stay.}
We used to think Intel would eventually remove the \texttt{clflush} instruction
due to the threat of flush-related attacks~\cite{Yarom2014, Gruss2016, Lipp2018, Kocher2019}.
To our surprise, Intel not only continues to support \texttt{clflush} in their new architectures
but also introduces new instructions with similar functionality, such as \texttt{clflushopt} and \texttt{CLWB}.
As described in the ISA Reference~\cite{Intel2015},
these instructions are added to reduce the performance overhead of accessing persistent memory~\cite{Rudoff2017}.
Since persistent memory is a promising memory technology gradually adopted by almost all major computer architectures,
cache flush instructions will remain in user land in the foreseeable future.
Even if their usage is limited to the privileged software,
prime-probe attacks from malicious kernels against other users/OSes~\cite{Zhang2016,Inci2016} or SGX enclaves~\cite{Haehnel2017}
are still practical.

\begin{figure}[bt]
\centering{
\includegraphics[width=0.37\textwidth]{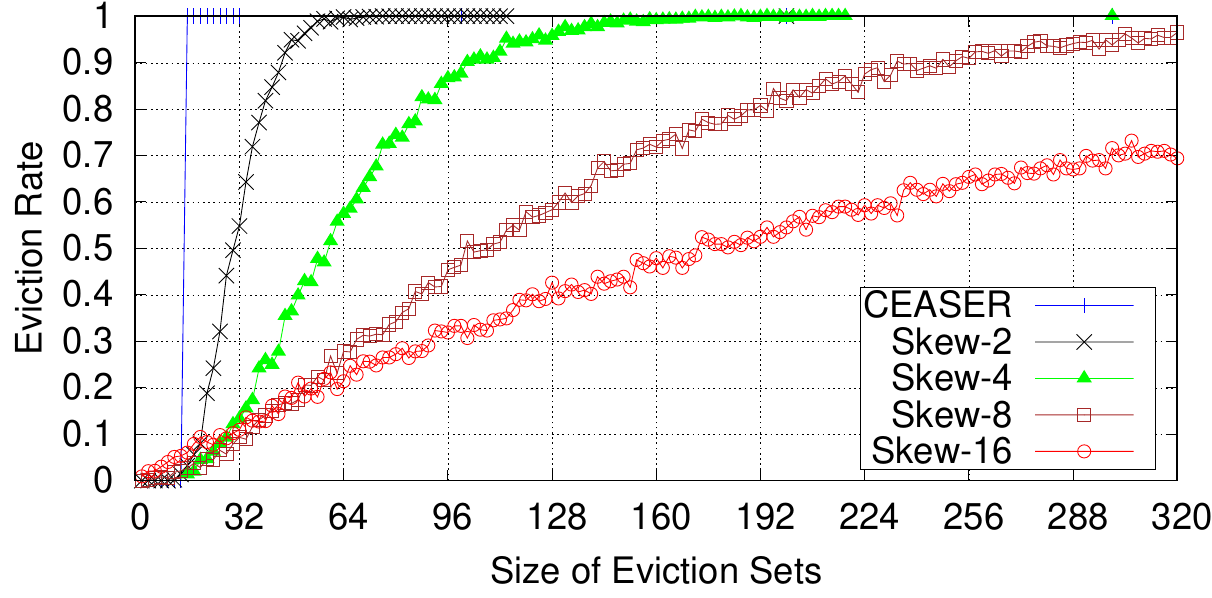}
}
\caption{
  The probability of evicting a target address (eviction rate) when a partially congruent eviction set is applied repeatedly.
  All LLCs are with the same size (1024-set 16-way).
  For skewed caches, the ways are divided into 2, 4, 8 and 16 partitions.
  Each result is averaged from 1000 independent experiments.
}
\label{fig:evset-effect}
\end{figure}

\begin{table}[bt]\footnotesize
  \centering
  \caption{Extracted from \figurename~\ref{fig:evset-effect},
    the estimated sizes of eviction sets to reach the expected eviction rates (30\%, 50\% and 80\%).}\label{tab:evset-size}
  \begin{tabular}{rccc}
    \toprule
    Cache Type        & 0.30 & 0.50 & 0.80 \\
    \midrule
    CEASER            & 16   & 16   & 16     \\
    Skew-2            & 25   & 30   & 39     \\
    Skew-4            & 45   & 59   & 87     \\
    Skew-8            & 68   & 108  & 190    \\
    Skew-16           & 90   & 172  & 400    \\
    \bottomrule
  \end{tabular}
\end{table}

Assuming attackers (can use cache flush instructions to) flush their eviction sets after each probe,
\figurename~\ref{fig:evset-effect} reveals the probability of evicting a target address (eviction rate)
when a partially congruent eviction set is applied repeatedly,
which complies with the theoretical analysis done in ScatterCache (\figurename~5 in \cite{Werner2019}):
the eviction rate increases with the size of the partially congruent eviction set.
\emph{When enough addresses are collected, a partially congruent eviction set can be used just like a fully congruent one.}

\textbf{Flawed hypotheses in ScatterCache}:
It is claimed in \cite{Werner2019} that attackers must find eviction sets with 99\% eviction rate
and must use a separate prime set to prime the LLC after each probe
(variant 1: single collision with eviction, Section~4.4~\cite{Werner2019}).
Both hypotheses are invalid.
\emph{An attacker can make use of eviction sets with low eviction rate} in persistent attacks.
Table~\ref{tab:evset-size} shows the number of partially congruent cache blocks
needed to achieve a certain eviction rate. 
An eviction set with a lower eviction rate is much smaller than a set with a higher eviction rate.
Since the time consumed in finding an eviction set is almost proportional to its size,
attacker can launch high-frequent attacks using small eviction sets to compensate the low eviction rate.
Even when a high eviction rate is required,
it can be achieved by repeatedly accessing an eviction set with a low eviction rate.
As describe in the flawed hypothesis in CEASER-S,
flushing the eviction set after each probe is much cleaner and faster than priming the LLC.

\subsection{Broken Defense}\label{sec:broken-defense}

\begin{figure}[bt]
\centering{
\subfloat[Conflict testing (CT)]
         {\includegraphics[angle=0, width=0.37\textwidth]{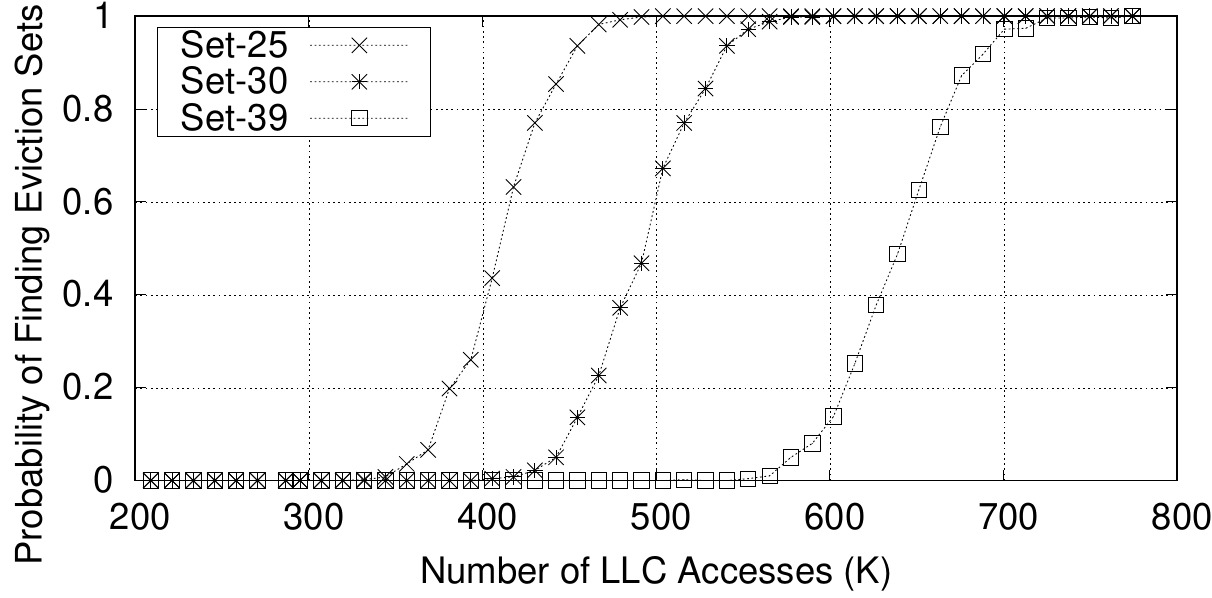}\label{fig:evset-attack}}\\
\subfloat[Prime, prune and then test (PPT)]
         {\includegraphics[angle=0, width=0.37\textwidth]{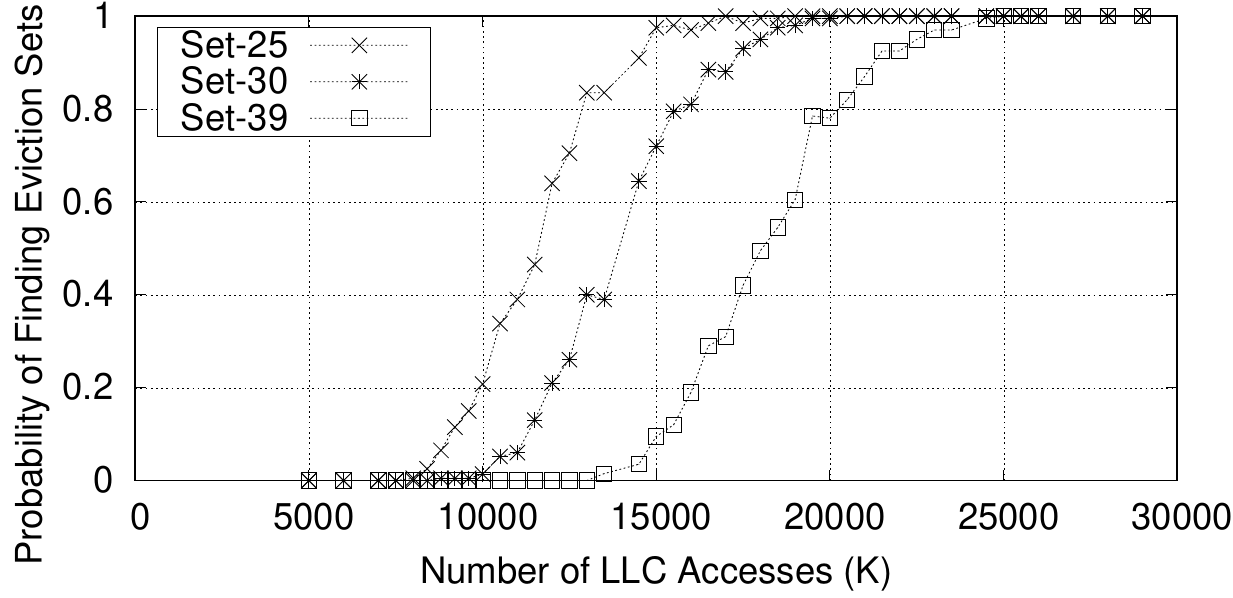}\label{fig:ppt-attack}}\\
}
\caption{
  The probability of finding eviction sets with 25, 30 and 39 partially congruent addresses
  in a skewed LLC (1024 sets, 16 ways, 2 patitions)
  within limited number of LLC accesses.
  Each result is averaged from 500 independent experiments.
}\label{fig:ceaser-s-attack-ana}
\end{figure}

In the three fast search algorithms, only CT and PPT
potentially work on randomized skewed caches.
Let us consider a CEASER-S LLC with two partitions~\cite{Qureshi2019}.
\figurename~\ref{fig:ceaser-s-attack-ana} demonstrates the probability of
finding eviction sets with 25, 30 and 39 partially congruent addresses
(corresponding to eviction rates of 30\%, 50\% and 90\% respectively)
using both CT and PPT.
As shown by the result,
although PPT is too long for any practical attacks (5M to 20M LLC accesses as shown in \figurename~\ref{fig:ppt-attack}),
it is possible to find a small eviction set (30\% eviction rate) in as low as 350K LLC accesses
using CT (\figurename~\ref{fig:evset-attack}),
which is far less than the preferred remap period of 1600K LLC accesses (100 accesses per cache block)~\cite{Qureshi2019}.
In fact, 1600K LLC accesses are long enough to find partially congruent eviction sets with 90\% eviction rate.

\begin{figure}[bt]
\centering{
\includegraphics[width=0.37\textwidth]{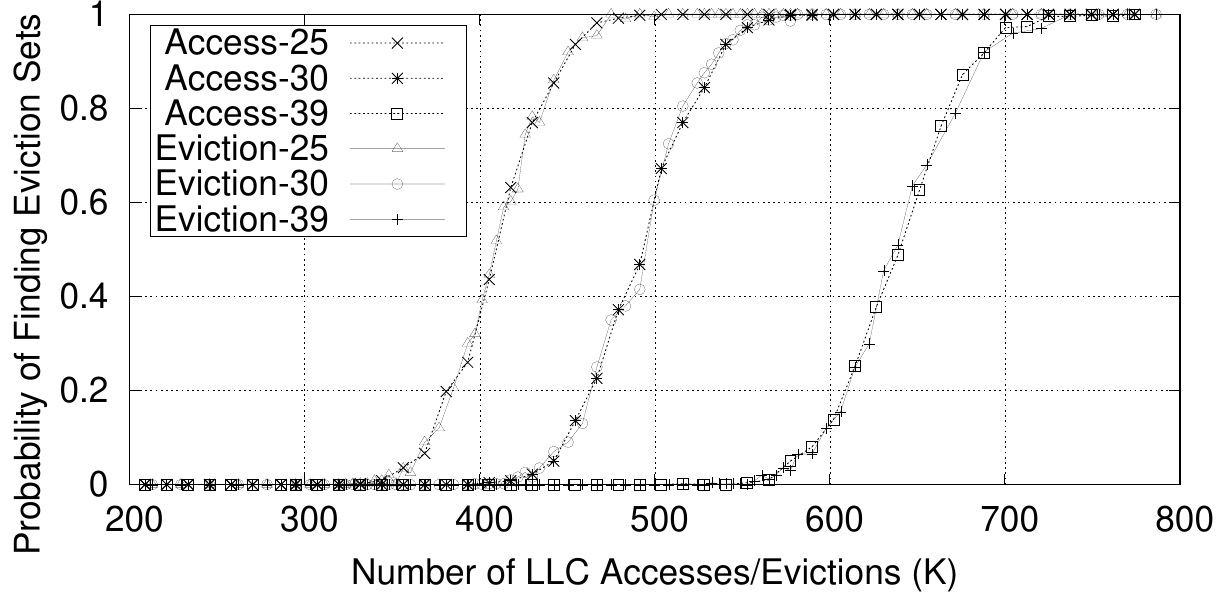}
}
\caption{
  The probability of finding eviction sets with 25, 30 and 39 addresses
  in a skewed LLC (1024 sets, 16 ways, 2 patitions)
  within limited number of LLC accesses or evictions.
  Each result is averaged from 500 independent experiments.
}
\label{fig:evset-attack-cmp}
\end{figure}

The reasons for the failure of CEASER-S are twofold:
One is its \emph{neglect of the possibility of using partially congruent eviction sets},
which require much lower number of LLC accesses to find than fully congruent eviction sets.
The other one is \emph{measuring the remap period by LLC accesses while overlooking the filter effect of private caches}.
As a separate test evaluating the probability of finding eviction sets within limited number of LLC evictions,
\figurename~\ref{fig:evset-attack-cmp} reveals that
nearly all accesses observed by the LLC are misses caused by random addresses.
According to the description in Section~\ref{sec:algorithm},
half of the cache accesses that re-access the target address are filtered by private caches.
The total number of cache accesses observed by the LLC is halved.

\begin{figure}[bt]
\centering{
\includegraphics[width=0.37\textwidth]{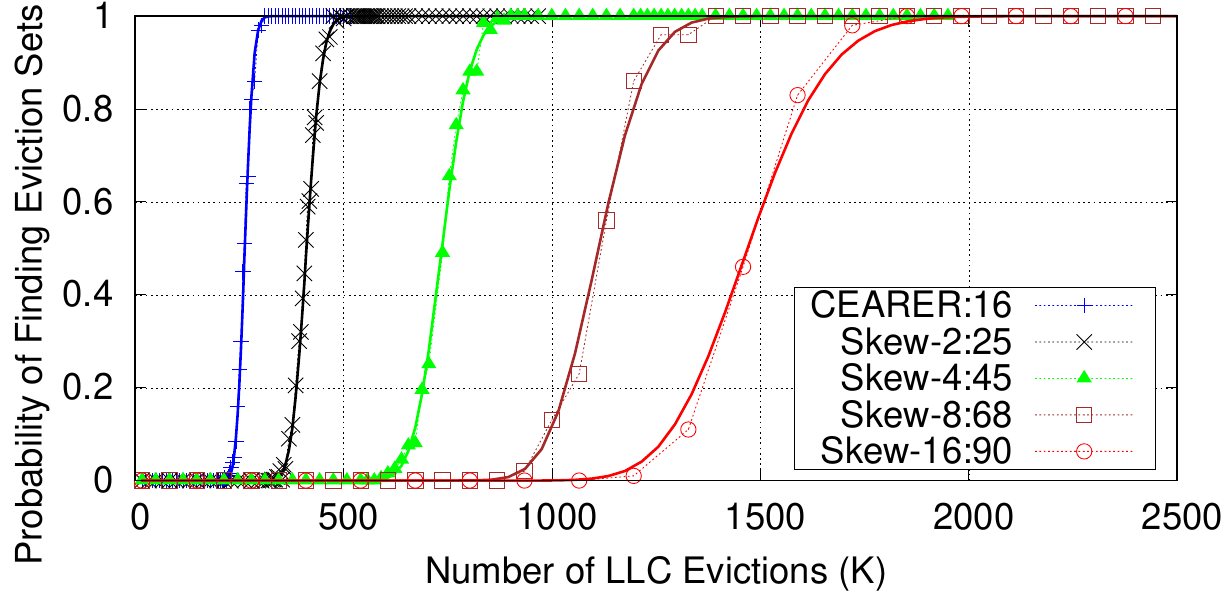}
}
\caption{
  The probability of finding the eviction sets with 30\% evict rate
  in various types of LLCs (CEASER and skewed cache with 2 to 16 partitions).
  Experimental results (averaged from 500 independent experiments) are depicted in dots while
  the probability in theory (Equation~\ref{eqn:succ-prob}) is drawn in solid lines. 
}
\label{fig:evset-attack-ev}
\end{figure}

Rather than periodically remapping the LLC,
ScatterCache proposes to use extra partitions to further increase the hardness in finding eviction sets
and assumes the extra hardness is enough to thwart attacks~\cite{Werner2019}.
ScatterCache estimates that roughly 275 partially congruent addresses are needed to
achieve the 99\% eviction rate in a randomized skewed cache with eight partitions
and finding such an eviction set
requires approximately
33.5M LLC evictions,
which is an intimidating large number.
\figurename~\ref{fig:evset-attack-ev} demonstrates the number of LLC evictions required
to finding a partially congruent eviction set with 30\% eviction rate
in all types of randomized caches.
If an attacker tries to find a small eviction set (68 addresses for 30\% eviction rate) instead of the large one,
the total number of LLC evictions is reduced to 1.1M,
which is only 3.3\% of what the large eviction set needs.\footnote{
  We believe that ScatterCache has over-estimated the number of victim accesses required.
  Instead of measuring the latency of re-accessing the random address,
  an attacker can measure the latency of accessing the target address (by the victim).
  This reduces the number of victim accesses to $n_{\text{\emph{ways}}} \cdot 2^{b_{\text{\emph{indices}}}} \cdot t$,
  which is $\frac{1}{n_{\text{\emph{ways}}}}$ of what ScatterCache estimates.
  }
Even if an attacker requires the 99\% eviction rate,
she can choose to re-access and flush the small eviction set 20 times.
The total number of LLC evictions is around 2.7K,
which is just a negligible fraction (0.2\%) of the LLC evictions needed for finding the small set.
\emph{The use of eviction sets with low eviction rate significantly speeds up attacks.}

\section{Fix the Randomized Skewed Caches}\label{sec:fix-skew}

Randomized skewed caches are still vulnerable to attacks
using partially congruent eviction sets found by the CT algorithm.
Several techniques are proposed in this section to strengthen the defense while retain performance.

\subsection{Count Cache Evictions Rather Than Accesses}

As analyzed in Section~\ref{sec:broken-defense},
the failure of CEASER-S is partially because the remap period is measured by LLC accesses
but half of the supposed accesses are filtered by private caches.
We propose to measure the remap period by LLC evictions.

In the CT algorithm,
when the target address is cached in the LLC,
the probability that a newly fetched random is cached in the same set and partition with the target address
can be described as:
\begin{equation}\label{eqn:}
P = \frac{1}{SK}
\end{equation}
where $K$ is the number of partitions.
Assuming the LRU replacement is used,
the target address is evicted from the LLC only when $\frac{W}{K}$
evictions occurred in the same set and partition,
which leads to $\frac{W}{K}$ evictions.
Therefore, the probability of collecting a partially congruent eviction set of $L$ addresses
in $E$ LLC evictions can be estimated as:
\begin{equation}\label{eqn:succ-prob}
Prob(X \ge L) = 1 - \sum_{i=0}^{\frac{LW}{K}-1}\binom{E}{i}P^i(1-P)^{E-i}
\end{equation}

As shown in \figurename~\ref{fig:evset-attack-ev},
the theoretical probability calculated using Equation~\ref{eqn:succ-prob}
matches with the experiment result.
We can use this equation to estimate the time of finding an eviction set (30\% eviction rate)
within different remap periods in various randomized caches.
Assuming the highest frequency of LLC evictions is 800 MHz,
Table~\ref{tab:time-estimate} details the time estimation.
If we consider one year as a secure time margin for thwarting potential attacks,
the chosen remap periods along with its time estimation are listed in the final column.
To safely thwart attacks,
the remap period of a two partitioned CEASER-S LLC must be reduced to 14 LLC evictions per cache block.
Even a skewed cache with 16 partitions has to be remapped very 39 LLC evictions per cache block.

\begin{table}[bt]\footnotesize
  \centering
  \caption{
    Estimated time for successfully finding an eviction set (30\% evict rate)
    within different remap periods (average number of evictions per cache block).
  }\label{tab:time-estimate}
  \begin{tabular}{lccccc}
    \toprule
    Cache Type        & 100   & 50    & 20      &  10      & Chosen Period\\
    \midrule
    CEASER            & 0.3ms & 0.3ms & 0.4ms   &  3.7y    & 10 (3.7y) \\
    Skew-2            & 0.5ms & 0.5ms & 0.32s   &  $>$100y & 14 (204y) \\
    Skew-4            & 0.9ms & 0.9ms & $>$100y &  $>$100y & 25 (40y) \\
    Skew-8            & 1.4ms & 2.8s  & $>$100y &  $>$100y & 35 (12y) \\
    Skew-16           & 1.8ms & 1.2h  & $>$100y &  $>$100y & 39 (12y) \\
    \bottomrule
  \end{tabular}
\end{table}

Such short remap periods might be considered intolerable.
However, \emph{remapping by counting LLC evictions is much more efficient than counting LLC accesses
because the LLC miss rate of normal applications is much lower than attacks.}
In an ideal scenario, if the miss rate in the LLC is sufficiently low,
remapping every 14 LLC evictions per cache block would trigger less remaps than
remapping every 100 LLC accesses per cache block (preferred by CEASER-S).
Our performance experiments in Section~\ref{sec:perf-norm} will analyze this effect in details.

\subsection{Multi-Step Cache Relocation}

Our experiment shows that 40\% to 50\% cache blocks in the LLC are evicted during the remap process,
which is why frequent remaps can hurt performance significantly.
Borrowing ideas from ZCache~\cite{Sanchez2010},
we propose to use a multi-step relocation in the remap process,
which reduces the eviction ratio to as low as 10\%.
This has two major benefits:
One is the reduced performance loss as extra blocks remain in the LLC.
The other one is the reduced damage from denial-of-service attacks~\cite{Kong2008}.
An attacker can trigger frequent remaps by forcing a large amount of LLC accesses or evictions.
Since the remap process cannot differentiate victim's data from the attacker's,
the attacker can use remaps as a stealthy way to blindly evict victim's data.
If the eviction ratio is reduced from 50\% to just 10\%,
the return of such attacks becomes marginal.

\begin{figure}[bt]
\centering{
\subfloat[]
         {\includegraphics[angle=0, width=0.12\textwidth]{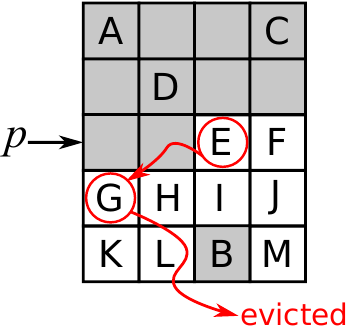}\label{fig:ceaser-relocate}}\\
\subfloat[]
         {\includegraphics[angle=0, width=0.32\textwidth]{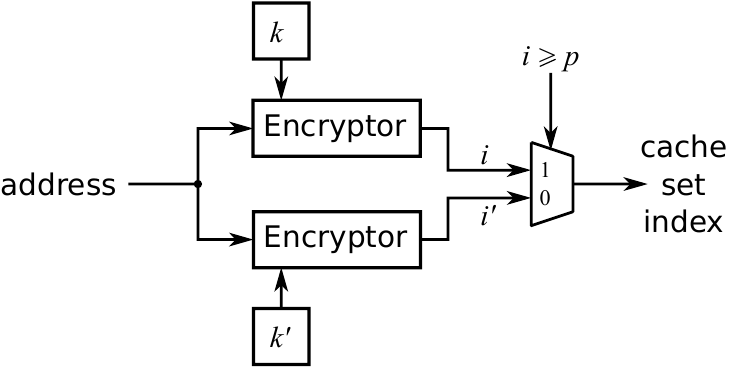}\label{fig:ceaser-key-select}}
}
\caption{
  The remap process of CEASER (CEASER-S).
  Cache sets are sequentially relocated as depicted in (a),
  where $p$ points to the cache set that s currently be relocated.
  During the remap process, the cache set index for an incoming address is decided accoridng to (b).
}\label{ceaser-remap}
\end{figure}

In the remap process proposed by CEASER~\cite{Qureshi2018},
cache sets are remapped sequentially as illustrated in \figurename~\ref{fig:ceaser-relocate}.
Remapped blocks are recorded in their metadata (shadowed in gray)
and a set-relocation pointer ($p$) always points to the cache set currently being remapped.
The cache block $E$ is currently being relocated to the next cache set chosen by the new key ($k'$).
According to the replacement policy, $G$ is evicted to make a room for $E$.
By repeating this procedure, all blocks in the set are remapped and $p$ moves to the next set.
Since remapping is a gradual procedure,
normal cache accesses might occur in parallel
and \figurename~\ref{fig:ceaser-key-select} illustrates how the cache set index is decided.
The old cache set index $i$ and the new one $i'$ are produced simultaneously
by two independent encrytors using the old key $k$ and the new key $k'$ respectively.
When $i \ge p$,
denoting the cache set is not remapped yet,
the old index $i$ is used.
Otherwise, the new index $i'$ should be used as the block is either remapped or missing.

The problem of the original remap process is the eviction of $G$.
Whenever the target cache set for a relocated block is full,
a cache block is evicted,
which leads to a high number of evictions at the beginning of the remap.

\begin{figure}[bt]
\centering{
\subfloat[]
         {\includegraphics[angle=0, width=0.105\textwidth]{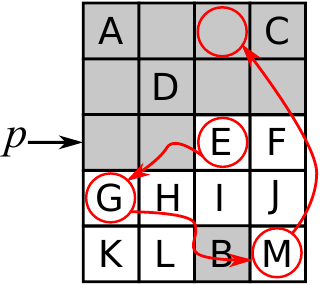}\label{fig:zcache-relocate}}\\
\subfloat[]
         {\includegraphics[angle=0, width=0.35\textwidth]{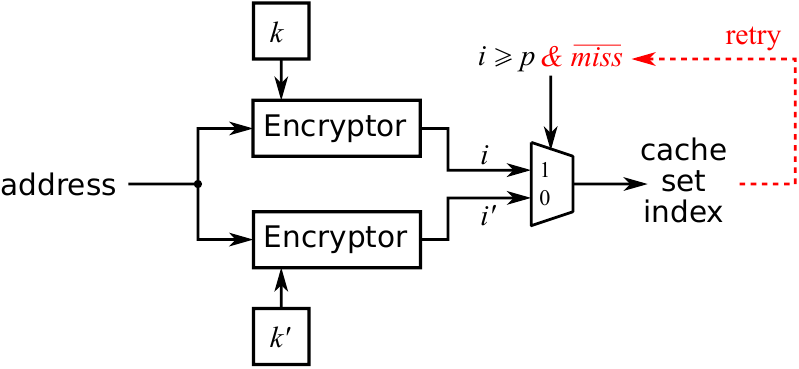}\label{fig:zcache-key-select}}
}
\caption{
  A multi-step relocation process.
  When the destination of a relocation is taken by an unremapped cache block,
  this block is further relocated until an empty space is found as in (a)
  or a remapped cache block is found and evicted instead.
  The cache set index for an incoming address is decided accoridng to (b).
  When using the old cache set index $i$ results in a miss, retry using the new index $i'$.
}\label{zcache-remap}
\end{figure}

\begin{figure}[bt]
\centering{
\includegraphics[width=0.37\textwidth]{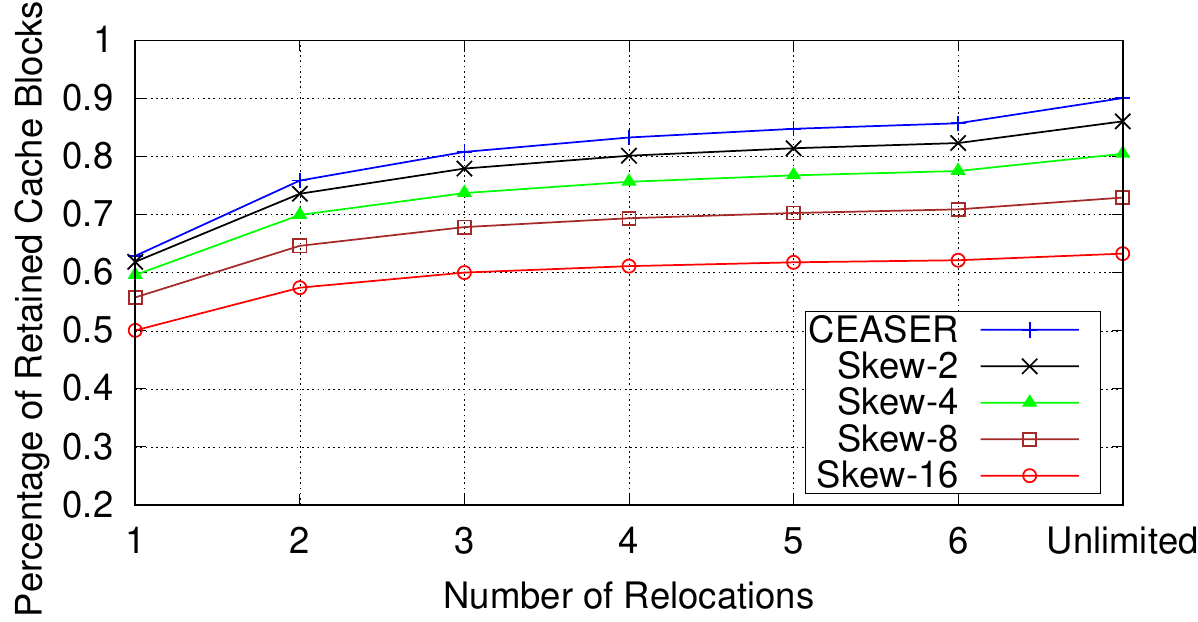}
}
\caption{
  The percentage of cache blocks retained during remapping
  by applying limited number of relocations.
  The maximum retaining percentage is achieved when `infinite' trials of relocation are applied
  until a remapped cache block is found as the replacement (and evicted).
  Each result is averaged from 100 independent experiments.
}
\label{fig:zcache}
\end{figure}

Such evictions might be avoidable.
The relocation procedure can keep on relocating the blocks to be evicted, such as $G$, in a chain
until either a free space if found, as shown in \figurename~\ref{fig:zcache-relocate},
or a remapped block is to be evicted.
Note that using multi-step relocation does not increases the total number of relocated blocks.
Once a block is relocated once, it is recorded as remapped and will not be relocated again.
The total number of blocks to be relocated is equal to the number of blocks in the LLC in both methods.
As shown in \figurename~\ref{fig:zcache}, by increasing the allowed number of relocations to unlimited,
the percentage of blocks retained in the LLC grows.
Randomized set-associative caches (CEASER) benefits the most
as the percentage increases from 63\% to 90\%.
The boost for randomized skewed caches drops gradually with the number of partitions.

The calculation of the cache set index needs a small change to support the multi-step relocation.
As shown in \figurename~\ref{fig:zcache-key-select},
when $i \ge p$,
the old cache set index $i$ should be used in the same way as in the original CEASER.
However, if it results in a miss,
the new cache set index $i'$ should be used in a retrial as
the block might have already be relocated.
Since reading the metadata array and checking cache hit typically finish in one or two cycles,
and retrials occur only occasionally during the relatively short remap process,
the performance impact is trivial considering the significantly reduced eviction ratio.

\begin{figure}[bt]
\centering{
\includegraphics[width=0.30\textwidth]{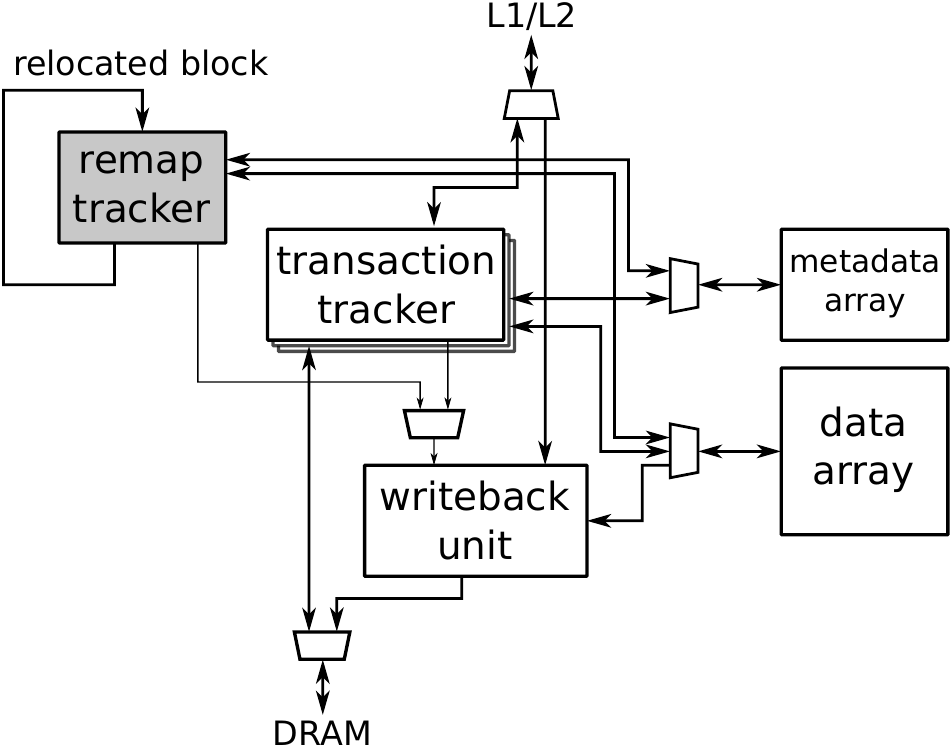}
}
\caption{
  Support multi-step relocation in the LLC of the Rocket-Chip.
}
\label{fig:remap-l2}
\end{figure}

Supporting multi-step relocation in the actual cache hardware should be straightforward.
\figurename~\ref{fig:remap-l2} demonstrates the internal structure of the LLC (L2) used in the Rocket-Chip SoC~\cite{Asanovic2016}
(available from lowRISC v0.4~\cite{lowRISC2017}),
which is a widely adopted open processor design taped out for tens of times.
To support multiple concurrent cache transactions initiated from the multiple L1 caches,
a cache slice implements multiple transaction trackers
sharing the same accesses to the metadata array, the data array, and the writeback unit.
When an incoming transaction is not blocked by race conditions,
a free tracker is allocated to serve it.
To support remaps, a special remap tracker is added.
During a remap, it tracks the set-relocation pointer and gradually relocates all cache blocks.
In the case of multi-step relocation,
when an unremapped cache block is swapped out,
the remap tracker throw it back to itself as a prioritized writeback transaction.
As long as unremapped blocks are swapped out, they are continuously relocated until
a free space is found or a remapped block is swapped out instead (which is evicted).
This recursive procedure effectively implements the unlimited steps of relocation.
The only hardware changes necessary to support multi-step relocation
are adding an incoming port to the remap tracker and modifying its state machine accordingly.

\section{Use Normal Caches Rather Than Skewed Caches}\label{sec:fix-ceaser}

Instead of advocating the use of randomized skewed cache like CEASER-S and ScatterCache,
we argue that randomized set-associative caches can be sufficiently strengthened and
possess a better chance to be actually adopted in commercial processors than their skewed counterparts.
Supported by a literature research with our best effort,
we cautiously believe that skewed caches~\cite{Seznec1993,Sardashti2014} have not yet adopted in LLCs of
any commercially available modern processors.
Promoting them purely for security benefits might be a hard sale.

\subsection{Issues with Skewed Caches}

We agree that skewed caches can improve cache efficiency by reducing conflicts~\cite{Seznec1993}
and are natural candidates for compressed caches~\cite{Sardashti2014}.
However, it seems that they are not yet embraced by the industry.
One potential reason has already been pointed out by CEASER-S~\cite{Qureshi2019}:
The benefit of skewed caches diminishes when the cache associativity increases.
As caches in modern processors are typically highly associative,
the marginal gain in performance might not justify the extra hardware cost.
For this reason, CEASER-S chooses to use only two partitions.
Some of our experiments show excessive skewing (too many partitions) actually hurt performance
as it reduces the efficiency of the LRU replacement.
One example is already revealed in \figurename~\ref{fig:zcache}.
The benefit of multi-step relocation drops with the increasing number of partitions.

From our own perspective in hardware designs, we also believe that
skewed cache significantly complicates the design of modern LLCs
which typically serve multiple cache transactions in parallel.
Taking the LLC design of the Rocket-Chip (as shown in \figurename~\ref{fig:remap-l2}) as an example,
before a tracker can accept a transaction,
the LLC must ensure that this transaction would not conflict with the others currently being served.
This typically means no two transactions served simultaneously should access the same cache set.
Otherwise, one of the conflicting transactions should be blocked before it is accepted by a tracker (race condition).
This is not a serious issue for set-associative caches
as the cache set index of an incoming transaction can be calculated beforehand
and compared with the indices of all active trackers simultaneously in a single cycle.
For a skewed cache with $K$ partitions and $T$ trackers,
the incoming transaction might access anyone of the $K$ possible cache sets
and it is not decided until it results in a hit or a target set is chosen for replacement.
In the worst scenario, $K \cdot TK$ parallel comparisons (rather than $T$ for the set-associative cache)
are required to check potential conflicts for an incoming transaction.
Besides the obvious hardware cost in doing so,
this significantly increases the
probability of blocking an incoming transaction due to a conflict that is not going to occur.
It then prolongs the cache accessing latency.

Therefore, we would like to investigate potential techniques
to strengthen the randomized set-associative caches.

\subsection{Remap When under Attack}

Although randomized skewed caches are vulnerable only to the CT algorithm,
randomized set-associative caches are vulnerable to all the three search algorithms introduced in Section~\ref{sec:algorithm}.
To thwart the CT algorithm, a 1024-set 16-way CEASER LLC
has to remap every 10 LLC evictions per cache block according to Table~\ref{tab:time-estimate},
which allows for a total of 160K LLC evictions between remaps.
However, our experiements show that
the numbers of LLC evictions (accesses) needed for finding an eviction set
are around 40.8K (168K) using the PPT algorithm
and 81.3K (532K) using the GE algorithm.
Both are valid threats.

\begin{figure}[bt]
\centering{
\subfloat[]
         {\includegraphics[angle=0, width=0.40\textwidth]{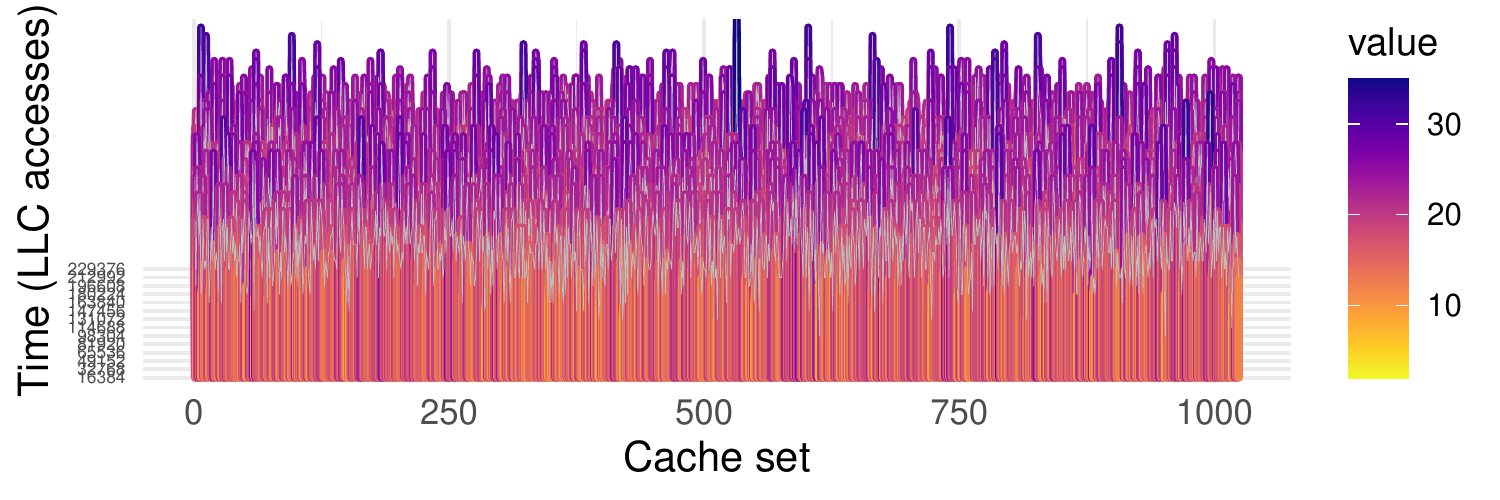}\label{fig:ppt-access}}\\
\subfloat[]
         {\includegraphics[angle=0, width=0.40\textwidth]{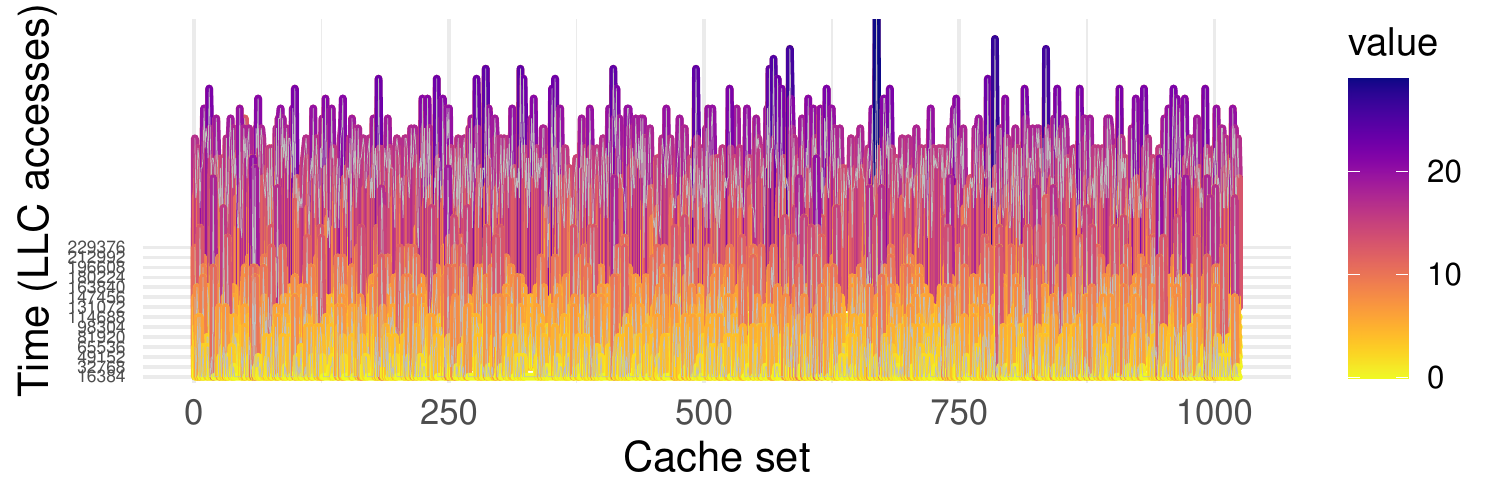}\label{fig:ppt-eviction}} \\
\subfloat[]
         {\includegraphics[angle=0, width=0.40\textwidth]{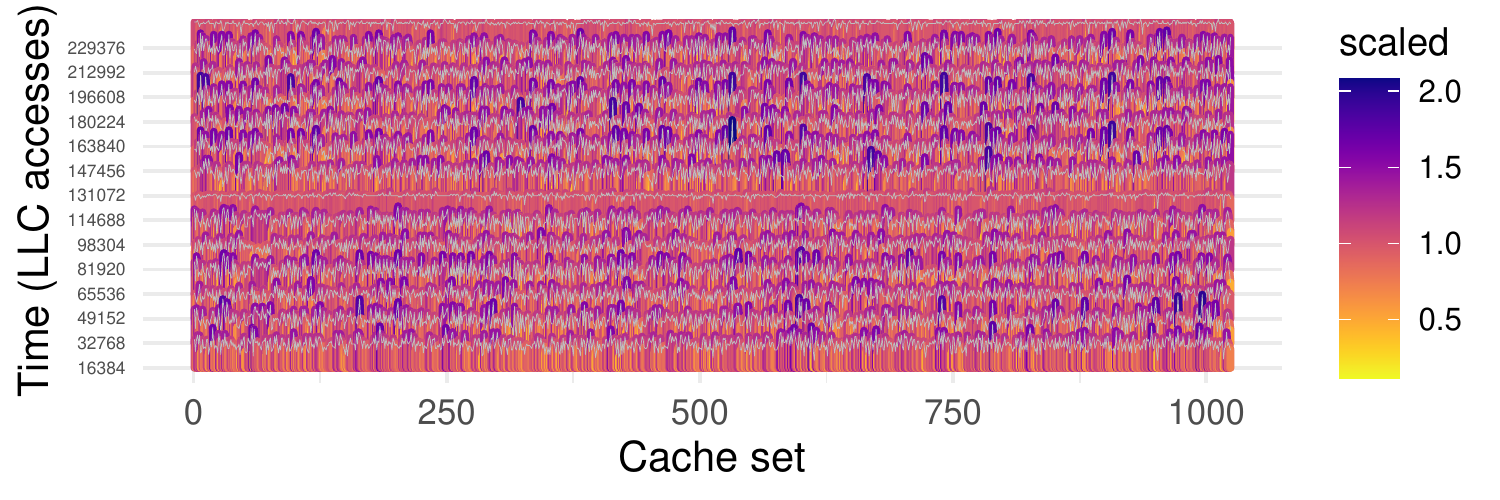}\label{fig:ppt-access-zscore}}\\
\subfloat[]
         {\includegraphics[angle=0, width=0.40\textwidth]{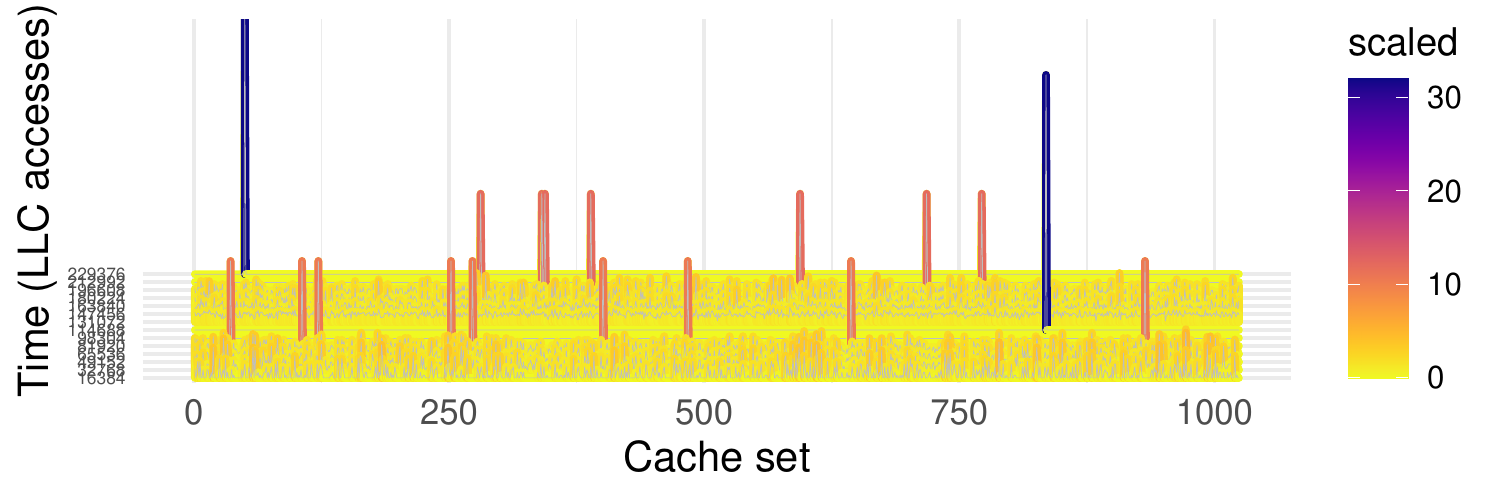}\label{fig:ppt-eviction-zscore}}
}
\caption{
  Detect the PPT attack by analyzing the cache set distributions of accesses and evictions.
  In all figures, the x-axis denotes the cache set index while
  the y-axis denotes simulation time measured in LLC accesses.
  The colored value denotes the number of accesses/evictions on a specific set occurred in a sample period.
  The chosen sample period is 16K LLC accesses.
  (a) and (b) depicts the set distribution of accesses and evictions during two round of attacks.
  (c) and (d) depicts the standardized version of (a) and (b) using the Z-Score method~\cite{Sugiyama2016, Scale}.
}
\end{figure}

Instead of shrinking the already short remap period,
we propose to trigger a remap when an attack using the two algorithms is detected
because both of them leave a unique pattern in the cache set distribution of evictions.
Let us first consider the PPT algorithm.
By periodically sampling the number of accesses and evictions occurred on individual cache sets during two consecutive attacks,
\figurename~\ref{fig:ppt-access} and \ref{fig:ppt-eviction} reveal the distribution of LLC accesses and evictions over all cache sets.
Both distributions seem totally random.
However, if we apply a Z-Score~\cite{Sugiyama2016} standardization on the distributions,
we can see two clear peaks in the standardized eviction distribution (\figurename~\ref{fig:ppt-eviction-zscore}),
although it is still random for the the standardized access distribution (\figurename~\ref{fig:ppt-access-zscore}).
The two peaks appear in the test phase of the PPT algorithm.
After the prune phase, all blocks in the prime set are concurrently cached in the LLC.
If there is any eviction in the test phase, it must occur on the target cache set.
As a result, the score of the target cache set reaches the maximum (32 for a 1024-set LLC) while it is zero on other sets.

\begin{figure}[bt]
\centering{
\includegraphics[width=0.42\textwidth]{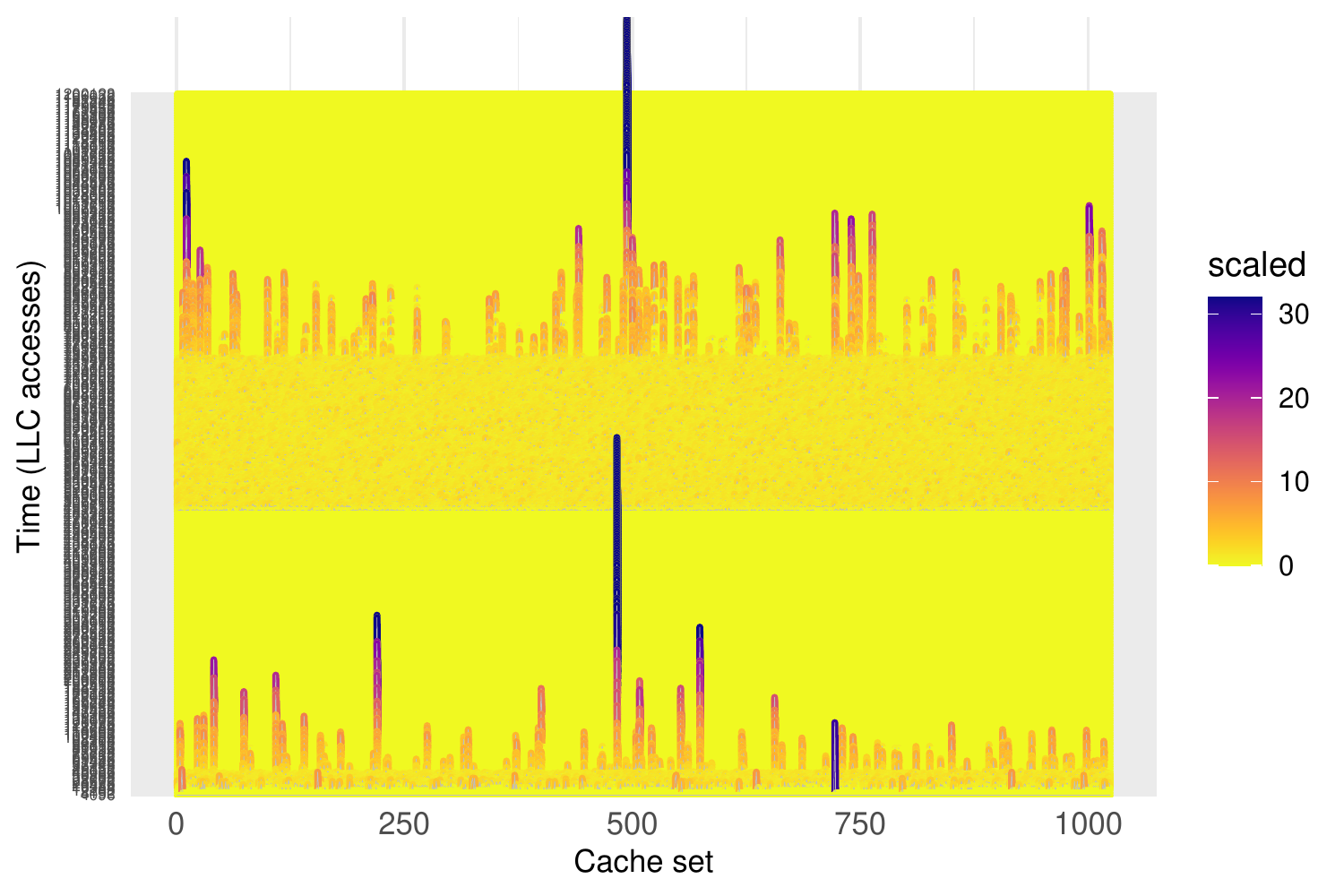}
}
\caption{
  Detect the \emph{group elimination} attack by analyzing the standardized cache set distribution of evictions.
}\label{fig:ge-attack-zscore}
\end{figure}

The GE algorithm presents a similar pattern
as demonstrated by \figurename~\ref{fig:ge-attack-zscore}.
Scores are small and randomly distributed at the early stage of the two simulated attacks
but converge on a single cache set when the large eviction set is finally condensed into a minimal one.

Since GE spends more time on condensing the eviction set than PPT testing the prime set,
detecting a PPT attack is harder than GE and we finalize our detector against the PPT algorithm (as it should work on GE as well).
We starts from a non-centered variant of the Z-Score standardization~\cite{Scale} to avoid negative scores:
\begin{equation}\label{eqn:zscore}
  z_i = \frac{e_i}{\sqrt{\frac{\sum{e^2}}{S-1}}}
\end{equation}
where $e_i$ is the number of evictions on cache set $i$ and $z_i$ is the calculated score for cache set $i$.
The score of the target cache set approaches to the maximum of $\sqrt{S}$ in an ideal attack.
However, reporting an attack whenever a maximum score is detected leads to false positive errors.
When the LLC miss rate is extremely low during normal operation,
there might be only one eviction during the whole sample period,
which also results in a maximum score.
To avoid such errors, we introduce the number of evictions into Equation~\ref{eqn:zscore} as weight:
\begin{equation}\label{eqn:weight-zscore}
  wz_i = (e_i - \bar{e}) \cdot z_i
\end{equation}
where $wz_i$ is the weighted score.
Sine an eviction set requires at least $W$ addresses,
the weighted score of the target cache set approaches to $W \cdot \sqrt{S}$ during the test phase of an ideal PPT attack.

An attacker can avoid detection if the detection threshold is simply set to $W \cdot \sqrt{S}$.
She can hide her trace by spreading the test phase over multiple sample periods.
In the extreme case, the attacker can collect only one congruent address in each round of PPT attack,
which effectively caps the weighted score to $\sqrt{S}$.\footnote{
  In practice, the number of rounds is limited because remaps will be triggered due to the excessive number of accumulated LLC evictions.
}
To detect such behavior and improve the robustness of the detector,
we apply an exponential moving average (EMA)~\cite{Hunter1986,Zhang2013} on the weighted score:
\begin{equation}
  az_i(t) = (1 - \alpha) \cdot az_i(t-1) + \alpha \cdot wz_i(t)
\end{equation}  
where $\alpha$ is a discount factor used to calculate $az_i(t)$, the EMA of $wz_i$ at sample $t$.
The use of EMA allows the detector to examine the history of $wz$
because $az$ is an infinite impulse response of $wz$.
$wz$ should be a zero-centered small number
for normal applications.
During the test phases of an attack, the $wz$ of the target cache set
unavoidably raises to at least $\sqrt{S}$.
By using a small $\alpha$,
the $az$ of the target cache set effectively accumulates the large $wz$ over the history,
which makes it sufficiently significant for detection.
We set the discount factor $\alpha$ to $\frac{1}{32}$ by a heuristic analysis.

\begin{figure}[bt]
\centering{
\includegraphics[width=0.33\textwidth]{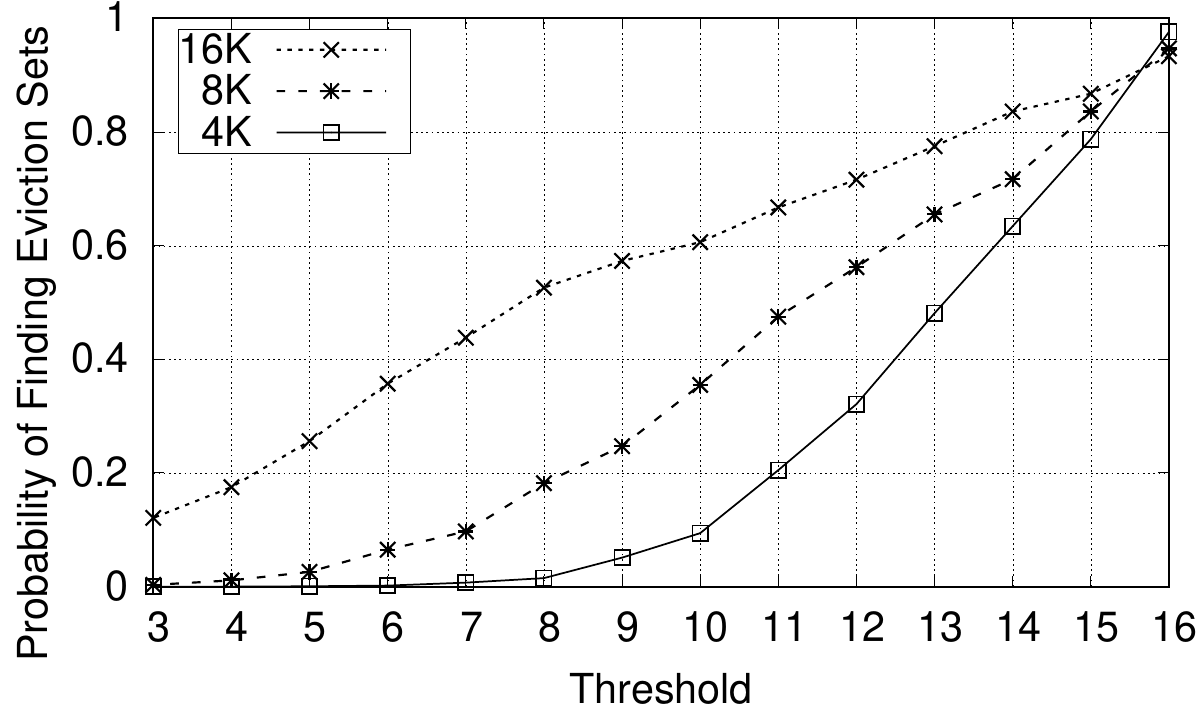}
}
\caption{
  The probability of finding an eviction set under active detection.
  Three sample periods are chosen: 16K, 8K and 4K LLC accesses.
  Each result is averaged from at least 200 independent experiments.
}
\label{fig:evset-create-rate}
\end{figure}

When the $az$ of a certain cache set reaches a threshold ($az \ge th$),
the detector triggers a remap.
The value of $th$ is crucial to the speed and the correctness of the detector.
The detector might leave a small window for a quick attack if a remap is late due to a large $th$.
However, normal programs might trigger remaps if $th$ is too small.
To choose a proper $th$, we run PPT attacks detected by different combinations of threshold and sample period (4K, 8K and 16K LLC accesses).
As shown in \figurename~\ref{fig:evset-create-rate},
sampling every 4K LLC accesses and triggering a remap whenever $az \ge 5$
is enough to reduce the probability of finding eviction set to almost nil.
Although not shown in the paper, we have verified that
GE attacks cannot escape detection with the same parameters.

\section{Performance}\label{sec:perf}

As the performance overhead of randomized caches has been
shown to be acceptable~\cite{Qureshi2018, Qureshi2019, Werner2019},
we analyze only the performance impact of the newly proposed techniques.

\subsection{Impact on Normal Applications}\label{sec:perf-norm}

\begin{figure}[bt]
\centering{
\includegraphics[angle=-90, width=0.47\textwidth]{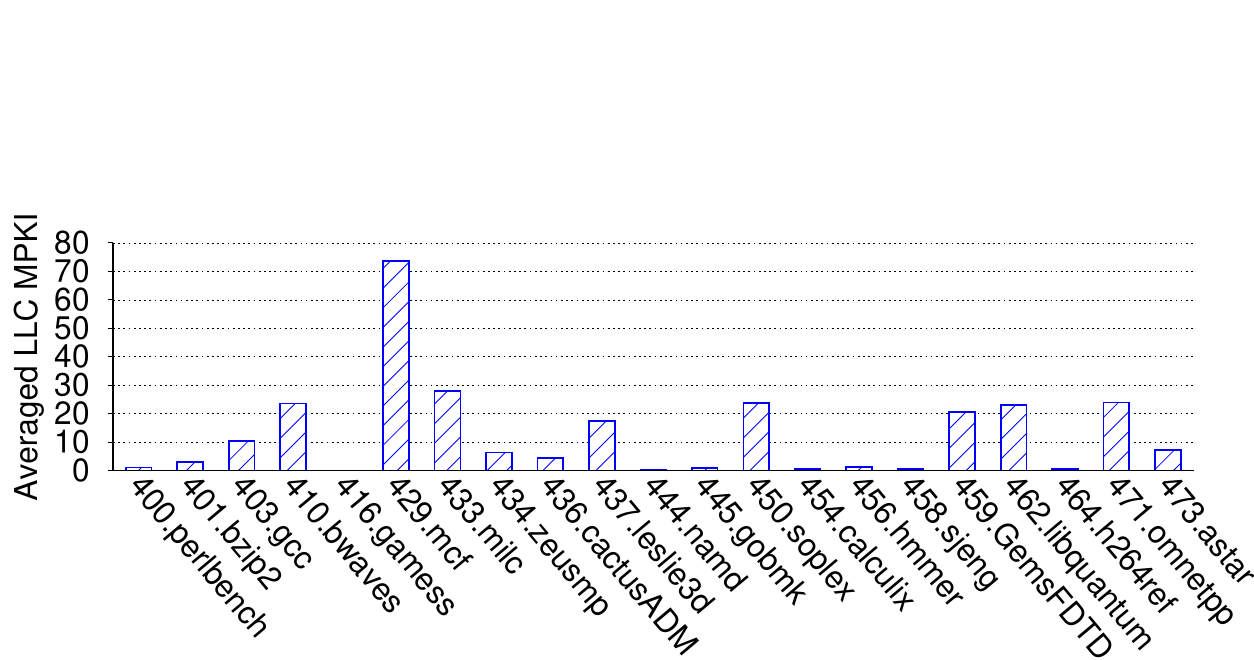}
}
\caption{
  MPKI of SPEC CPU 2006 benchmark cases using a static 1024-set 16-way CEASE LLC.
}
\label{fig:spec-mpki-ceaser-static}
\end{figure}

The SPEC CPU 2006 benchmark suite~\cite{Henning2006} is used to evaluate the impact on normal applications.
Similar to ScatterCache, performance results are measured without concurrent processes~\cite{Werner2019}.
As described in Section~\ref{sec:ana-platform},
we use a modified Spike simulator~\cite{Spike} as the evaluation platform.
A processing core has two private L1 data and instruction caches (16KB, 64-set, 8-way, 64B cache block).
A 1024-set 16-way L2 cache is used as the LLC where all randomized caches are implemented.
All cache levels use the LRU replacement.
Thanks to the fast simulation speed of Spike,
we are able to run 100G instructions for each benchmark case,
which is 100 and 400 times of the instructions simulated in CEASER\cite{Qureshi2018} and ScatterCache~\cite{Werner2019}.
\figurename~\ref{fig:spec-mpki-ceaser-static} shows the number of misses per K instructions (MPKI) using a static CEASE LLC,
which is used as the baseline for other performance results.
The MPKI figures match with the result provided in CEASER\cite{Qureshi2018}.

\begin{figure}[bt]
\centering{
\includegraphics[angle=-90, width=0.47\textwidth]{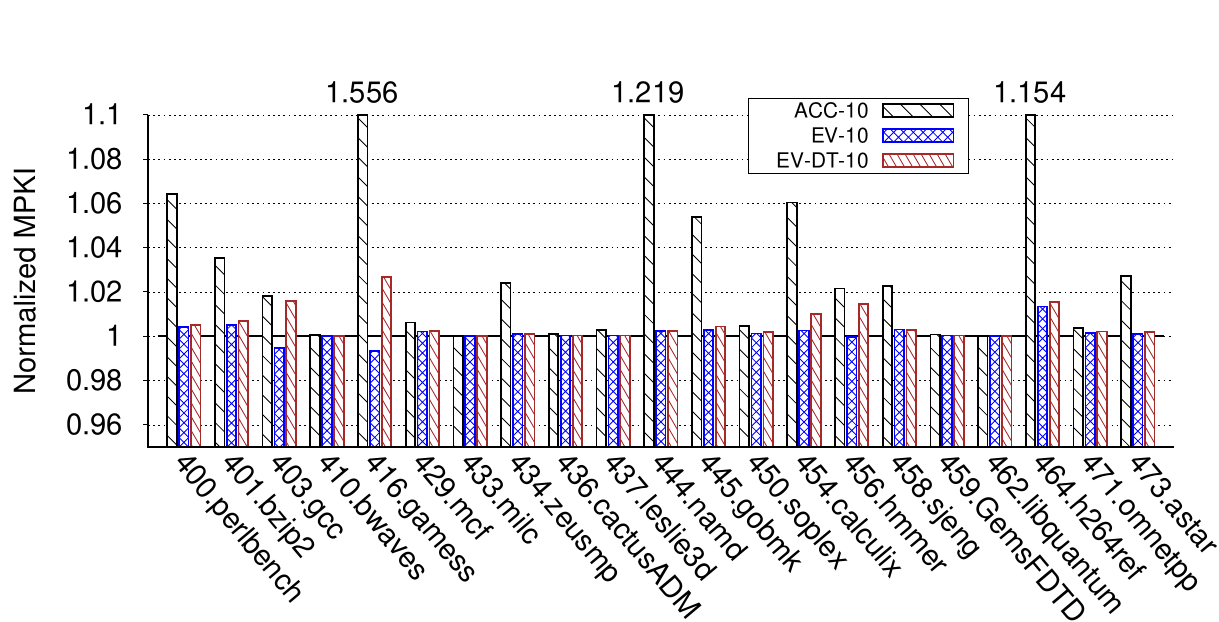}
}
\caption{
  Normalized MPKI of SPEC CPU 2006 benchmark cases running on a CEASER LLC.
  ACC-10: remap every 10 accesses per cache block;
  EV-10: remap every 10 evictions per cache block;
  DT: attack detection.
  The static CEASE is used as the baseline.
}
\label{fig:spec-ceaser-mpki}
\end{figure}

\figurename~\ref{fig:spec-ceaser-mpki} demonstrates the performance overhead of different remap strategies on a CEASER LLC.
The remap period is increased to 10 accesses/evictions per cache block to thwart the CT attack.
The average overhead is 0.61\%, 0.077\% and 0.19\% for
ACC-10 (remapping by accesses), EV-10 (remapping by evictions) and EV-DT-10 (EV-10 plus attack detection) respectively.
Measuring the remap period by evictions rather than accesses reduces MPKI by 69\% with attack detection or 87\% without. 

\begin{figure}[bt]
\centering{
\subfloat[Remaps per G instructions (RPGI)]
         {\includegraphics[angle=-90, width=0.47\textwidth]{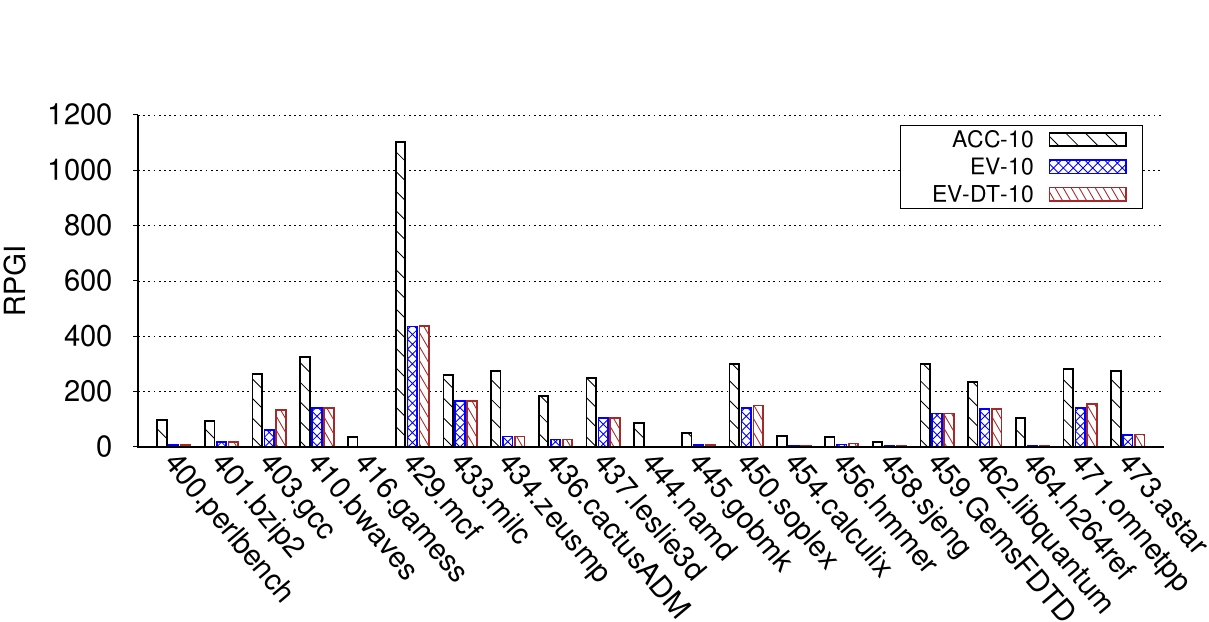}\label{fig:spec-ceaser-remap-ab}}\\
\subfloat[Normalized RPGI using ACC-10 as the baseline]
         {\includegraphics[angle=-90, width=0.47\textwidth]{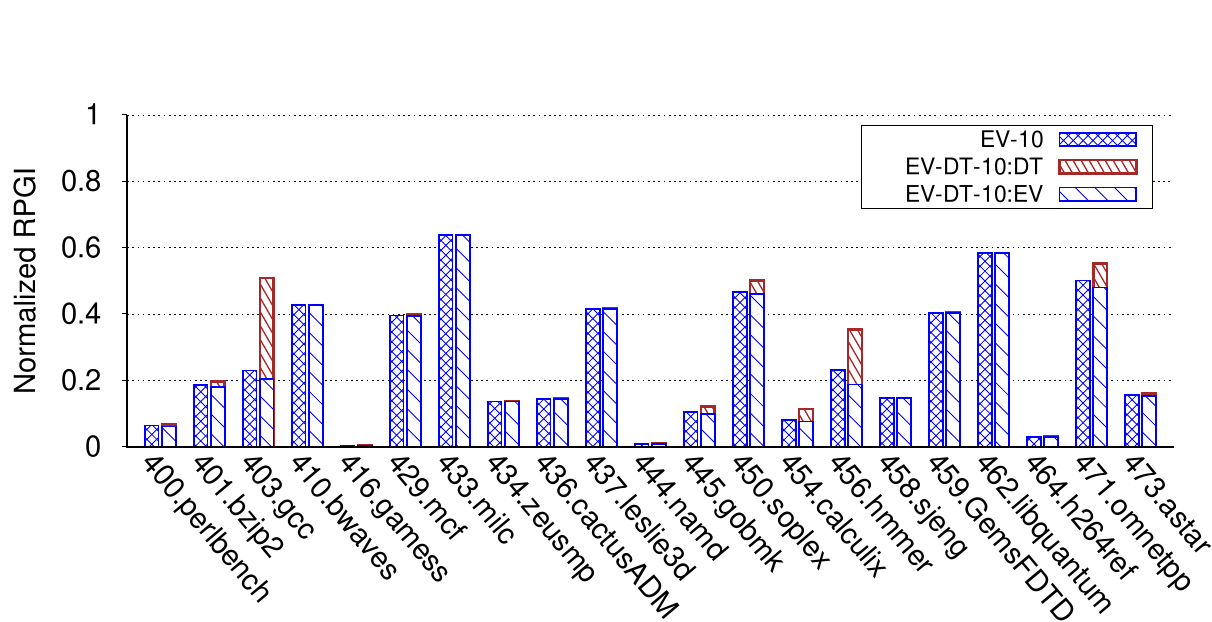}\label{fig:spec-ceaser-remap-norm}} \\
}
\caption{
  Remaps per G instructions (RPGI) of SPEC CPU 2006 benchmark cases.
  In (b), DT and EV denote the remaps triggered by attack detection and reaching remap period respectively. 
}
\label{fig:spec-ceaser-remap}
\end{figure}

\begin{figure}[bt]
\centering{
\subfloat[MPKI (remap period: 100 accesses per cache line)]
         {\includegraphics[angle=0, width=0.47\textwidth]{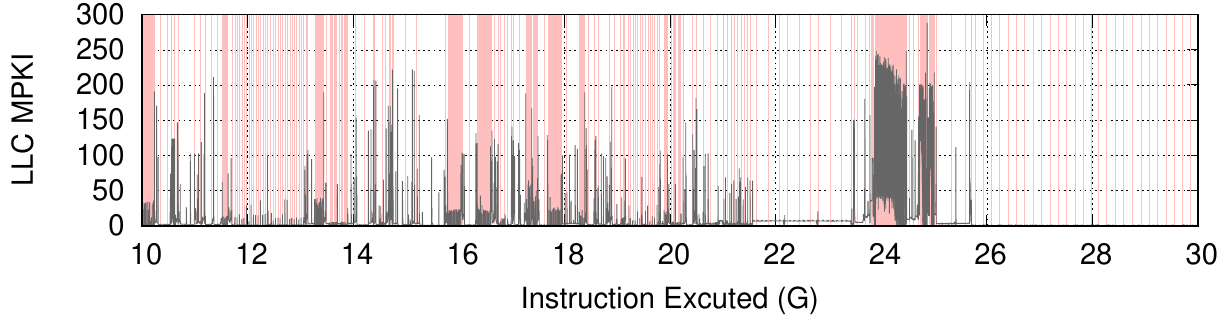}\label{fig:mpki-org}}\\
\subfloat[Miss rate (remap period: 100 accesses per cache line)]
         {\includegraphics[angle=0, width=0.47\textwidth]{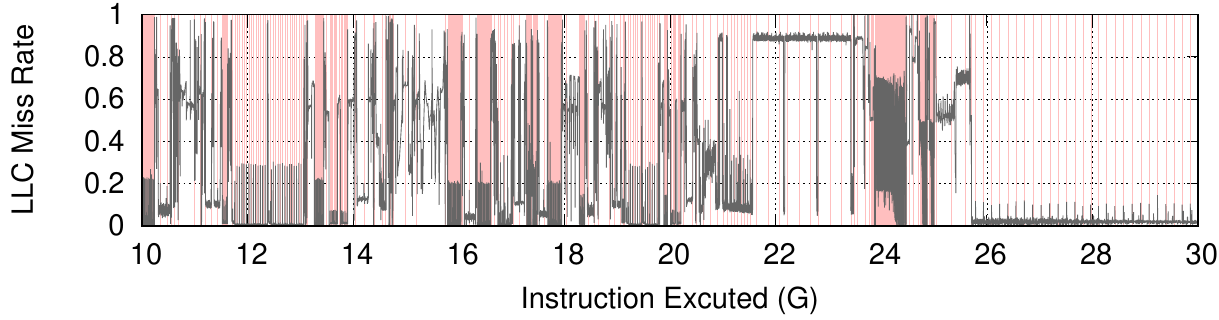}\label{fig:mrate-org}} \\
\subfloat[MPKI (remap period: 10 evictions per cache line)]
         {\includegraphics[angle=0, width=0.47\textwidth]{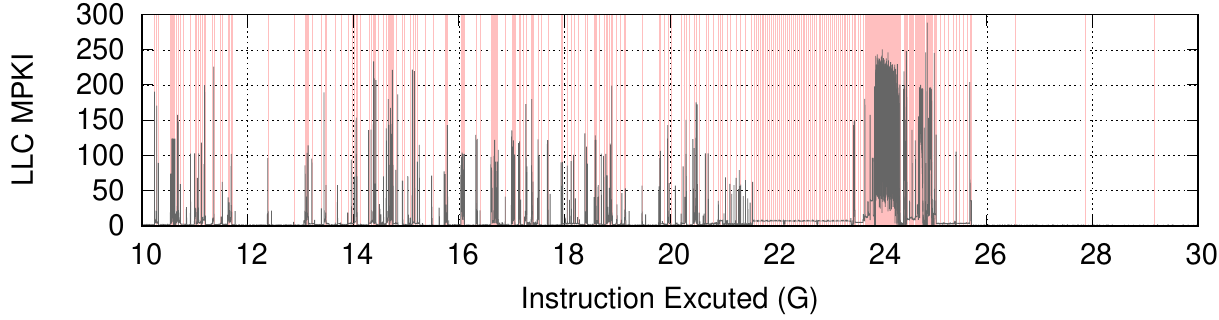}\label{fig:mpki-ev}}\\
\subfloat[Miss rate (remap period: 10 evictions per cache line)]
         {\includegraphics[angle=0, width=0.47\textwidth]{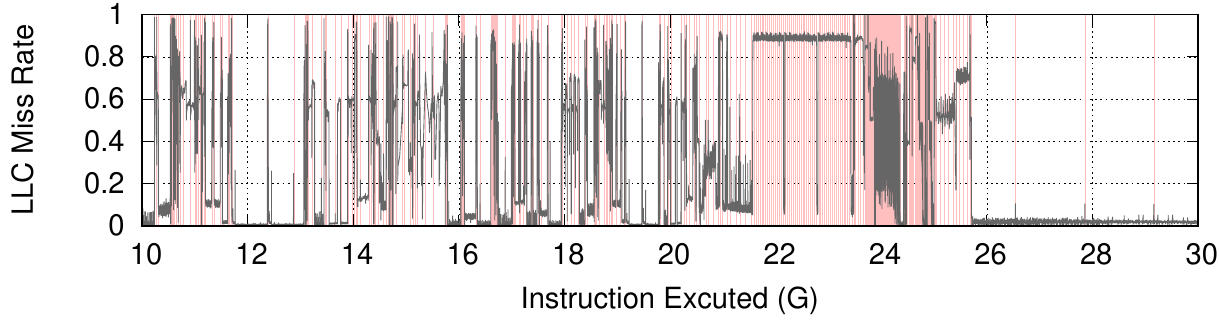}\label{fig:mrate-ev}}
}
\caption{
  Compare the triggered remaps of running the SPEC CPU 2006 case 403.gcc (expr2).
  Each remap is depicted as a vertical pink line in the background.
}\label{fig:reduced-remap-impact}
\end{figure}

Such significant performance boost comes from two reasons:
One is the reduced number of remaps as shown in \figurename~\ref{fig:spec-ceaser-remap-ab}.
The average reduction is 64\% with attack detection or 71\% without.
The other one is the reduced impact for each remap.
To explain this effect,
\figurename~\ref{fig:reduced-remap-impact} depicts the run-time MPKI and miss rate curves
extracted from a representative window of the 403.gcc (expr2) benchmark case.
Note that for the LLC remapped by accesses,
the remap period is increased to 100 accesses per cache block to avoid excessive remaps (a totally pink colored background).
Remapping by accesses inclines to remap when both MPKI and miss rate are low,
such as the time segments of (11--13), (19--20), and (26--30) G instructions,
while there is nearly no remaps when remapping by evictions.
What is worse, these remaps lead to unnecessary block evictions which in turn raise the miss rate.
On the contrary, remapping by evictions inclines to remap when the miss rate is high,
such as the time segments of (22--23) and (25--26).
During these segments, the utilization efficiency of the LLC is already reduced by the high miss rate.
The performance impact of the unnecessarily evicted blocks in each remap is thus weakened.

The cost of enabling attack detection in CEASER is relatively small compared with the performance boost from remapping by evictions.
As shown in \figurename~\ref{fig:spec-ceaser-remap-ab}, the cost of detection in only 7\% of the original cost of remapping by accesses.
\figurename~\ref{fig:spec-ceaser-remap-norm} provides a detailed analysis of the remaps triggered by detection.
For most benchmark cases, the number of mistakenly detected attacks (false positive errors) is tiny.
Only cases like 403.gcc and 456.hmmer have high numbers of false positive errors.
Since the absolute number of remaps for 456.hmmer is extremely low ($\text{RPGI} \approx 8$ in \figurename~\ref{fig:spec-ceaser-remap-ab}),
the high rate of false positive errors does not actually hurt performance.
As for 403.gcc, the absolute number of MPKI increased from 10.25 to 10.47, leading to a 2.1\% increase.
Considering the MPKI is relatively low, a 2.1\% increase on the the low MPKI should be tolerable.

\begin{figure}[bt]
\centering{
\includegraphics[width=0.35\textwidth]{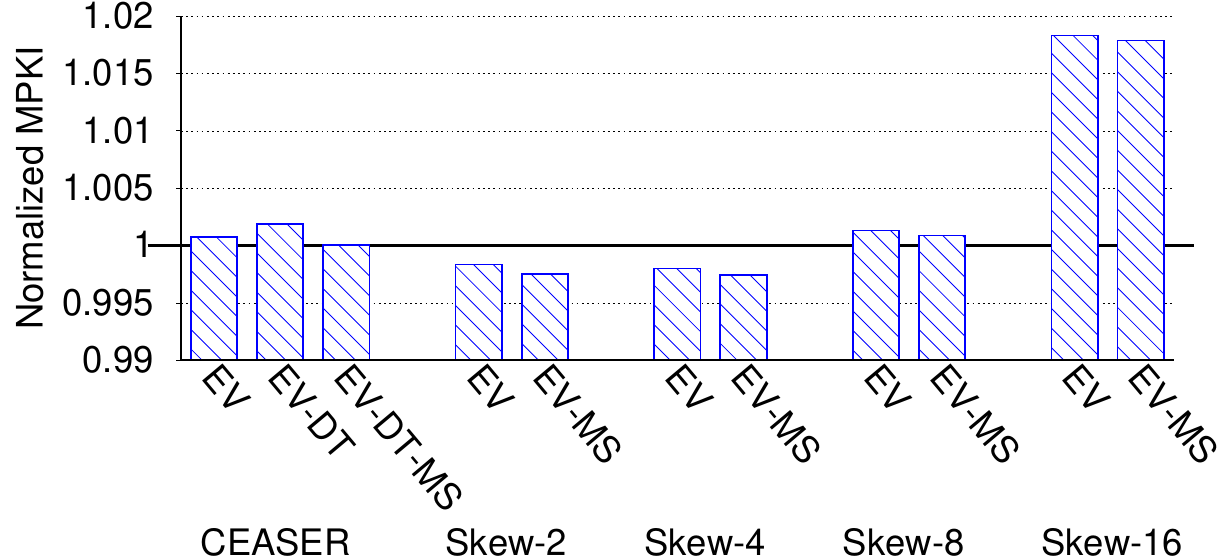}
}
\caption{
  Normalized MPKI of SPEC CPU 2006 benchmark cases using the static CEASE as the baseline.
  EV: remapping by evictions;
  DT: attack detection;
  MS: multi-step relocation.
}
\label{fig:spec-all}
\end{figure}

\figurename~\ref{fig:spec-all} shows the normalized MPKI of all types of randomized caches using the static CEASE as the baseline.
In general, skewed caches with a moderate number of partitions indeed reduce MPKI but such reduction is marginal (less than 0.5\%).
When more than eight partitions are used, MPKI begins to rise and introduce performance loss.
This is why we blieve randomized set-associate caches (CEASER) should be used if they are safely strengthened.
For CEASER LLCs, periodically remapping by evictions introduces 0.08\% extra MPKI
and enabling attack detection adds another 0.11\%,
but adopting the multi-step relocation would reduce the overhead back to a trivial 0.007\%.
This result shows that we can make the randomized set-associate cache safe enough without significant performance loss.
As for skewed caches,
utilizing the multi-step relocation reduces MPKI roughly by 0.05\% and the skewed cache with only two partitions benefits the most (0.08\%).
This complies with our estimation in \figurename~\ref{fig:zcache}.

\subsection{Logic and Memory Overhead}

The memory overhead of randomized caches has been analyzed in \cite{Qureshi2018, Qureshi2019};
therefore, we estimate only the extra cost using the new techniques.
We use a single core Rocket-Chip implemented by lowRISC (version 0.4)~\cite{lowRISC2017} as the base.
Using the same configuration as used in the Spike simulator,
the LLC (L2 cache) consumes around 22\% logic and 99\% SRAM of the processor (without outer AXI buses and devices).
To support remaps, a remap tracker is added to the LLC
which original has two acquire (access) trackers and one release (writeback) tracker.
The extra area overhead would be round 7.6\% logic of the processor (34\% logic of the LLC).
This overhead is relatively high but unavoidable.
Remapping by evictions rather than accesses introduces no area overhead.
The overhead of supporting multi-step relocation is also marginal
because the only changes required are adding a port to the remap tracker and modifying its state machine.
To estimate the overhead of attack detection,
we made a prototype of the detector in hardware.
The hardware detector finishes each round of detection in 2K cycles (less than the sample period of 4K LLC accesses).
By shrinking the precision of the intermediate results and reducing multiplier/divider to adder/shifter,
the detection error is within 5\% compared with the software implementation while
the area overhead (after place and route) is around 0.8\% logic and 0.4\% SRAM of the processor (3.5\% logic and 0.4\% SRAM of the LLC),
both of which are marginal.

\section{Discussion}\label{sec:related}

\textbf{New cache designs}:
Since the introduction of randomized skewed caches,
two new designs~\cite{Ramkrishnan2019,Tan2020} have been proposed and both of them promote the use of set-associative caches.
Indirection table (iTable) based two level dynamic randomization (TLDR)~\cite{Ramkrishnan2019}
tries to strengthen CEASER by another layer of randomization using an iTable.
An address is first randomly mapped to an iTable entry and
then the entry is mapped to a random cache set.
It is claimed that the extra iTable provides higher level of randomness than randomized skewed caches
and gradually remapping iTable entries reduces the remap-related performance loss.
PhantomCache~\cite{Tan2020} proposes to place an incoming cache block in one of the randomly selected cache sets
rather than partitions as in skewed caches.
This increases the level of randomness and allows the use of LRU for the whole cache set.
Both designs can safely defeat the GE attack but their effectiveness against CT and PPT attacks needs further investigation.
Finally Doblas~\cite{Doblas2020} extends the cache randomization from LLC to the L1 caches by using simple randomization functions.

\textbf{Performance evaluation}:
The performance results of all existing caceh randomization designs
come from various Gem5 simulations~\cite{Binkert2011, Qureshi2018, Qureshi2019, Werner2019, Tan2020},
whose slowness
limited the total number of instructions that can be simulated in resaonable time,
which further contrains the coverage on respective workloads~\cite{Sherwood2002, Nair2008}.
Our choices of using the fast (event-driven and timeless) Spike simulation
allows us to boost the number of simulated instructions by $100 \sim 400$ times,
which signifcantly increases the coverage on respective workloads,
but limits the performance evaluation in miss rate only,
leaveing the overhead on CPU execution time unstated.
We believe this is a reasonable trade-off as
the estimation on CPU execution time is inaccurate and cannot be used to compare between designs
even if the slowest Gem5 OoO model~\cite{Binkert2011} is used.
The reason is the lack of consensus on which encryption algorithm should be adopted
especially after the one used by CEASER has been found problematic~\cite{Bodduna2020}.
It is still an open challenge to choose a strong and fast encryption algorithm for randomized caches.
As a result, we do not have a fair way to evaluate the execution time
and cache miss rate is the only frequently used and unbiased metric available.

\textbf{Attack detection}:
Run-time detection of cache side-channel attacks
using the existing performance counters (pfc)~\cite{Zhang2013, Chiappetta2016, Payer2016, Zhang2016b, Chen2014}
has shown to be effective to detect persistent attacks by software.
Some software detectors adopt machine learning algorithms to increases the detection accuracy~\cite{Chiappetta2016, Zhang2016b}
but they are always constrained by the limited information available from pfc.
Hardware detectors~\cite{Yao2019, Harris2019} begin to appear recently.
Most of them exploit the cyclic pattern between an attacker and her victim~\cite{Chen2014, Yao2019, Harris2019}.
The concentration of cache accesses on certain cache sets during the exploitation phase
has long been discovered~\cite{Zhang2013, Oren2015, Fuchs2015}.
The exponential moving average was also used in software detectors~\cite{Zhang2013}.
Nevertheless, to the best of our knowledge,
we are the first to use a simple hardware detector to detect the searching of eviction sets
utilizing the concentrated set distribution.

\textbf{New attacks}:
Purnal improves the original PPT attack~\cite{Qureshi2019} by introducing the prune phase
and points out it is possible to use partially congruent eviction set to launch covert channel attacks on ScatterCache~\cite{Purnal2019}.
Our simulation and analysis on PPT are based on Purnal's work
but with our own optimized prune method as it is not clearly described in \cite{Purnal2019}.
Our experiments show that PPT attacks would fail on randomized skewed caches
because the accumulated number of LLC evictions always surpasses the proposed remap period.
However, it is very likely that the prune process can be further optimized to reduce its footprint in evictions.

\section{Conclusion}\label{sec:con}
We have newly discovered several problems with the hypotheses and implementations in the latest randomized skewed caches:
The possibility of using cache flush instructions in conflict-based attacks has been overlooked.
The concept of minimal eviction set no longer applies to randomized skewed caches.
Attackers do not have to use eviction sets with 99\% eviction rate.
Measuring the remap period by LLC accesses is flawed.
As a result, existing randomized skewed caches are still vulnerable to conflict-based cache side-channel attacks.

We proposed several defense techniques/suggestions to fix the newly discovered problems:
Measure the remap period by LLC evictions rather than accesses while further reduce the period.
Adopte ZCache-like multi-step relocation to minimize the number of cache blocks evicted during the remap process.
Our experiments show that all the newly discovered vulnerabilities are fixed within the current performance budget.
We also claim that randomized set-associative cache can be sufficiently strengthened with reasonable overhead
using a simple attack detection mechanism.
Compared with randomized skewed caches,
randomized set-associative caches are better candidates for future commercial processors.

\bibliographystyle{IEEEtran}
\bibliography{IEEEabrv,reference}

\end{document}